\documentclass[aps,prl,twocolumn,preprintnumbers,linenumber,superscriptaddress,amsmath,amssymb]{revtex4-1}
\usepackage{graphicx}
\usepackage{subfigure}
\usepackage{epsfig}
\usepackage{dcolumn}
\usepackage{bm}
\usepackage{dcolumn}
\usepackage{color}
\usepackage{amsbsy}
\usepackage{lineno}
\usepackage{blindtext}

\def\avg#1{\left\langle#1\right\rangle}
\def\bra#1{\left\langle#1\right|}
\def\ket#1{\left|#1\right\rangle}

\def\be{\begin{equation}}       \def\ee{\end{equation}}
\def\bea{\begin{eqnarray}}      \def\eea{\end{eqnarray}}
\def\ba{\begin{array} }
\def\ea{\end{array} }
\def\bnum{\begin{enumerate} }
\def\enum{\end{enumerate}}

\def\nn{\nonumber}
\def\pa{\partial}
\def\=>{\Rightarrow}
\def\>{\rightarrow}
\def\A{\uparrow}
\def\V{\downarrow}

\def\eye2{Fathbb{I}}

\def\Eq#1{Eq.~(\ref{#1})}
\def\Fig#1{Fig.~\ref{#1}}

\newcommand{\Ref}[1]{Ref.~\cite{#1}}

\renewcommand{\>}{\rangle}

\begin{document}
\title{Edge quantum criticality and emergent supersymmetry in topological phases}
\author{Zi-Xiang Li}
\affiliation{Institute for Advanced Study, Tsinghua University, Beijing 100084, China}
\author{Yi-Fan Jiang}
\affiliation{Institute for Advanced Study, Tsinghua University, Beijing 100084, China}
\author{Hong Yao}
\email{yaohong@tsinghua.edu.cn}
\affiliation{Institute for Advanced Study, Tsinghua University, Beijing 100084, China}
\affiliation{State Key Laboratory of Low Dimensional Quantum Physics, Tsinghua University, Beijing 100084, China}
\affiliation{Collaborative Innovation Center of Quantum Matter, Beijing 100084, China}

\begin{abstract}
Proposed as a fundamental symmetry describing our Universe, spacetime supersymmetry (SUSY)  has not been discovered yet in nature. Nonetheless, it has been predicted that SUSY may emerge in low-energy physics of quantum materials such as topological superconductors and Weyl semimetals.
Here, by performing state-of-the-art sign-problem-free quantum Monte Carlo simulations of an interacting two-dimensional topological superconductor, we show convincing evidence that the $\mathcal{N}$=1 SUSY emerges at its edge quantum critical point (EQCP) while its bulk remains gapped and topologically nontrivial. Remarkably, near the EQCP, we find that the edge Majorana fermion acquires a mass that is identical with that of its bosonic superpartner. To the best of our knowledge, this is the first observation that fermions and bosons have equal dynamically generated masses, a hallmark of emergent SUSY. We further discuss experimental signatures of such EQCP and associated SUSY.
\end{abstract}
\date{\today}
\maketitle

As a spacetime symmetry interchanging fermions and bosons, supersymmetry (SUSY) was originally introduced in particle physics to attempt to solve various fundamental issues such as the hierarchy problem \cite{Weinbergbook, wessbook, Gervais-Sakita1971, Wess-Zumino1974, Dimopoulos-Georgi1981}. In a theory with unbroken SUSY, each particle and its superpartner would share the {\it same} mass and most internal quantum numbers except spin. As the present Universe is apparently not supersymmetric, many experiments including the recent ones at the LHC have been trying to look for evidence of SUSY at higher energy or spontaneous SUSY breaking; but no definitive results have been found so far \cite{LHCbook,LHCfootnote}.

The question ``Under what circumstances can SUSY emerge at low energy and long distance in quantum many-body systems that are not supersymmetric at lattice scale?'' has attracted increasing attention. Remarkably, it has been shown theoretically that SUSY may emerge at low energy in certain condensed matter systems at quantum criticality \cite{Shenker-84, foda1988, Balents1998, Fendley2003, SSLee2007, Luijse2008, Grosfeld2011, Bauer2013, KYang2010, Ashvin2014, Ponte2014, Berg2015, Jian2015,  Franz2015, zerf2016, Jian2016, Grover2016, Subirbook}, mainly from renormalization group (RG) analysis of the low-energy effective theory at the corresponding quantum critical (multicritical) points where the terms breaking SUSY are irrelevant in the infrared limit. Especially, it was shown theoretically in Ref. \cite{Ashvin2014} that SUSY can emerge at the quantum critical point on the boundary of topological superconductors (TSC) \cite{Hasan-Kane-10,Qi-Zhang-11} by tuning a single parameter while the bulk is still fully gapped and topologically nontrivial. However, {\it edge quantum critical points} (EQCP) in microscopic models of two-dimensional interacting topological superconductors \cite{comment} have not been unambiguously revealed so far. Moreover, to investigate emergent SUSY at the putative EQCP, we need to accurately determine critical exponents of the quantum phase transitions.
Consequently, a two-dimensional microscopic model of interacting topological superconductors and its {\it unbiased} solution demonstrating such EQCP and emergent SUSY are vastly desired not only for their own interest but also for future experimental verifications of emergent SUSY.

\begin{figure}[t]
\includegraphics[width=8.6cm]{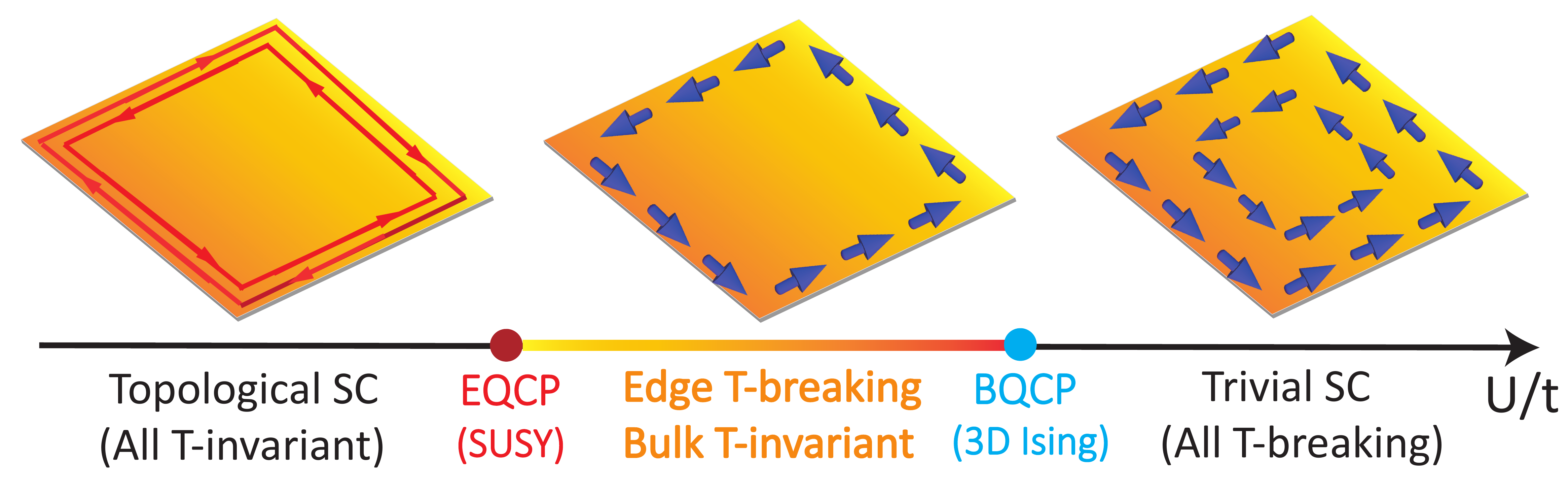}				
\caption{The quantum phase diagram of an interacting two-dimensional topological superconductor (SC) revealed by the sign-problem-free QMC simulations. Here, the EQCP represents an ``edge quantum critical point'' where time-reversal breaking with magnetic order occurs only on the edge but not in the bulk. At the EQCP, spacetime SUSY emerges by tuning only a {\it single} parameter, the Hubbard interaction $U$. Similarly, BQCP labels the ``bulk quantum critical point'', beyond which both the edge and bulk break time-reversal symmetry. }
\label{schematic}
\end{figure}

Here, we introduce a minimal model of interacting time-reversal-invariant TSC of spin-1/2 electrons on the square lattice [see \Eq{tsc} below] to fill in this gap. Importantly, this minimal model is sign-problem free in determinant quantum Monte Carlo (QMC) simulations \cite{BSS1981, Hirsch1985, Loh1990, Troyer2005, Zaanen2008, Assaad2008} by employing time-reversal symmetry \cite{Congjun2005,Li2015}. Recently, a number of sign-problem-free algorithms and related fermionic systems were studied \cite{Li2015,Berg2012,Lei-2015,Berg2016prx, Shailesh-14,Li2016,Xiang2016,Assaad2016, Ashvin2017NatPhys,Li2016SciBull,Berg2016sdw, Kivelson2016arxiv,Li2015arxiv,Meng2016arxiv}. By performing state-of-the-art Majorana QMC simulations of this interacting model of topological $p+ip$ superconductors, we find that with increasing interactions the topological superconductor's helical edge Majorana fermions first undergo spontaneous time-reversal breaking while its bulk remains gapped and topologically nontrivial (bulk time-reversal breaking occurs only at stronger interaction strength). The quantum phase diagram is shown in \Fig{schematic}. More importantly, the critical exponents at the EQCP obtained by our large-scale QMC simulations are consistent with exact results of the $\mathcal{N}$=1 supersymmetric theory, which provides convincing evidence that the EQCP in helical Majorana edge states of a TSC features an emergent spacetime SUSY \cite{Ashvin2014}. To the best of our knowledge, this is the {\it first} time that an EQCP with emergent spacetime SUSY has been observed by intrinsically unbiased simulations of a two-dimensional quantum many-body model.

Emergent SUSY at quantum criticality dictates that fermions and their bosonic superpartners have equal masses at and near the QCP. However, so far such dynamically generated phenomena have not been observed in condensed matter systems. In the Letter, by computing the imaginary-time Green's functions by large-scale QMC simulations, we are able to compute the masses of the edge Majorana fermion and its bosonic superpartner near the EQCP. We find that they {\it equal} to each other within a numerical error bar: namely $m_f$=$m_b$ where $m_f$ and $m_b$ are fermion and boson masses, respectively. This is the first observation that fermions and bosons have equal dynamically generated masses, a hallmark of emergent SUSY. We emphasize that the equal masses between fermions and bosons observed near the EQCP are emerge phenomena, rather than an explicit assumption in microscopic models \cite{KYang2010}. When the system moves sufficiently away from the EQCP, the fermion mass gradually differs from the boson mass, due to the explicit breaking of SUSY away from the EQCP. Such phenomena shed light to our understanding of the possible breaking of assumed SUSY in nature. \\

{\bf Sign-problem-free model of interacting TSC:} We first introduce a minimal model describing interacting TSC of spin-1/2 electrons on the square lattice with time-reversal symmetry ($\mathrm{T}$):
\bea\label{tsc}
H \!=\! \sum_{ij,\sigma} \!\big[\!-t_{ij} c^\dagger_{i\sigma} c_{j\sigma} \!+\!\Delta_{ij,\sigma}c^\dagger_{i\sigma}c_{j\sigma}^\dagger \!+\! h.c.\!\big]\! \!-\!U\sum_{i} n_{i\uparrow} n_{i\downarrow},~~~
\eea
where $c^\dag_{i\sigma}$ creates an electron on site $i$ with spin polarization $\sigma$=$\uparrow$,$\downarrow$,  $n_{i\sigma}$=$c^\dag_{i\sigma}c_{i\sigma}$ is the number operator, $t_{ij}$=$t$ labels nearest-neighbor hopping, and $t_{ii}$=$\mu$ is the chemical potential. Hereafter we set $t=1$ as the unit of energy. Here $\Delta_{ij,\A}$=$ \Delta(\delta_{j,i\pm\hat x}+i\delta_{j,i\pm \hat y})$ and $\Delta_{ij,\V}$=$ \Delta(\delta_{j,i\pm\hat x}-i\delta_{j,i\pm \hat y})$ such that the Hamiltonian in \Eq{tsc} with a finite $\Delta$ describes a time-reversal invariant TSC \cite{Schnyder-08, Kitaev-09, Qi-Raghu-09} with ($p$+$ip$) triplet pairing of spin-up electrons and ($p-$$ip$) pairing of spin-down electrons.

For weak Hubbard $U$ interaction, the system's edges host massless helical Majorana fermions that are protected by the $\mathrm{\it T}$: $T$=$i\sigma^yK$, where $\sigma^i$ acts in spin space and $K$ represents complex conjugation. Note that, besides the $\mathrm{T}$, the model also respects another symmetry $P$=$\sigma^z$; namely, the spin parity $(-1)^{N_\A}$ is conserved where $N_\A$ is the total number of spin-up electrons. The topological classification of interacting superconductors respecting both symmetries was studied in Refs. \cite{Qi-13,Yao-13,Ryu-12}. More importantly, simulating the minimal model in \Eq{tsc} with attractive Hubbard interactions is {\it sign-problem-free} 
(more specifically it belongs to the sign-problem free Majorana class \cite{Li2015,Li2016,Xiang2016}). 
Consequently, the correlation effect in such TSC can be investigated by large-scale Majorana QMC simulations. In particular, we investigate whether an EQCP and emergent SUSY at the EQCP can be realized by tuning a single parameter, {\it i.e.}, the interaction $U$. \\

\begin{figure}[t]
\subfigure[]{\includegraphics[height=3.0cm]{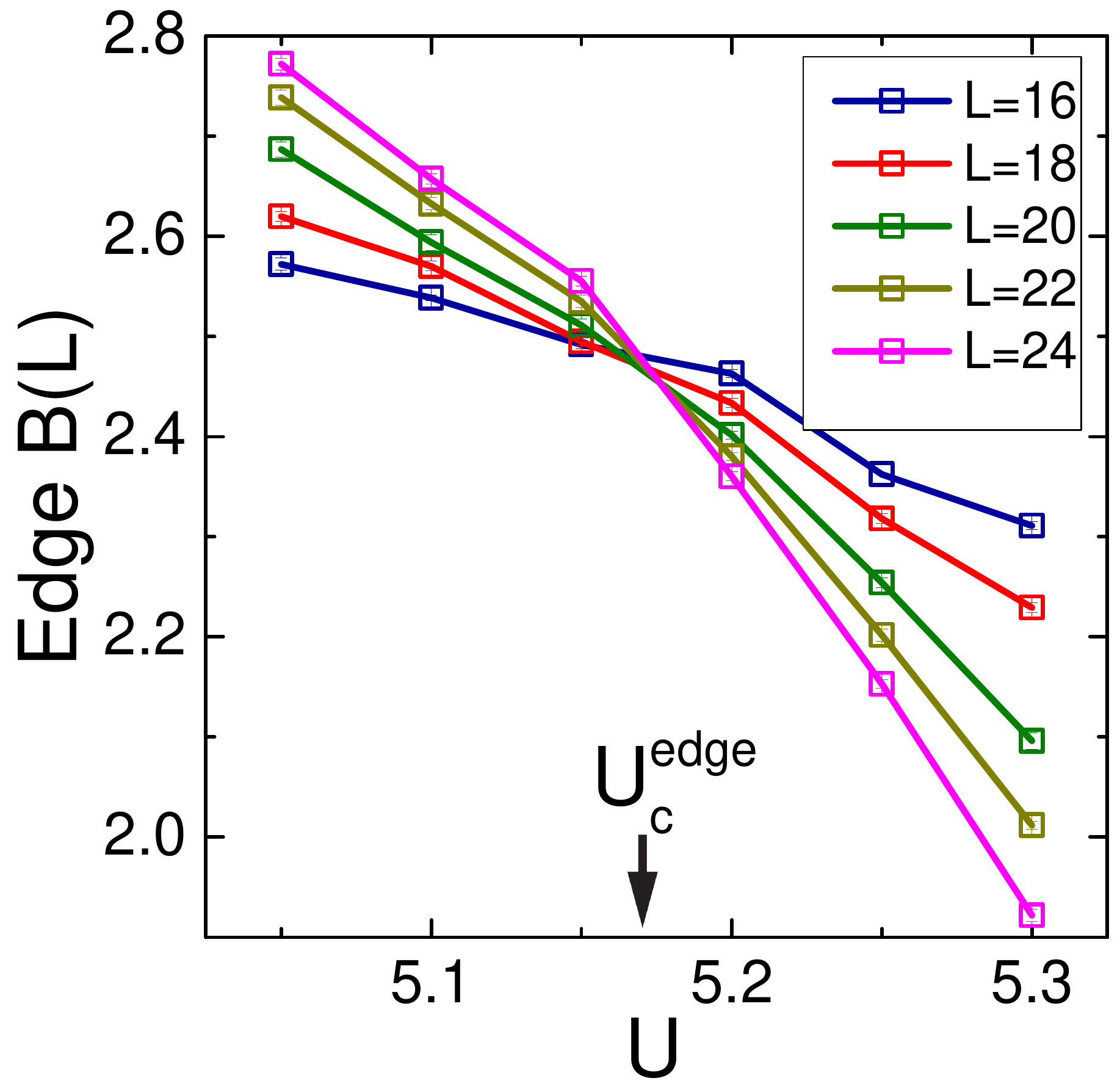}}
\subfigure[]{\includegraphics[height=3.0cm]{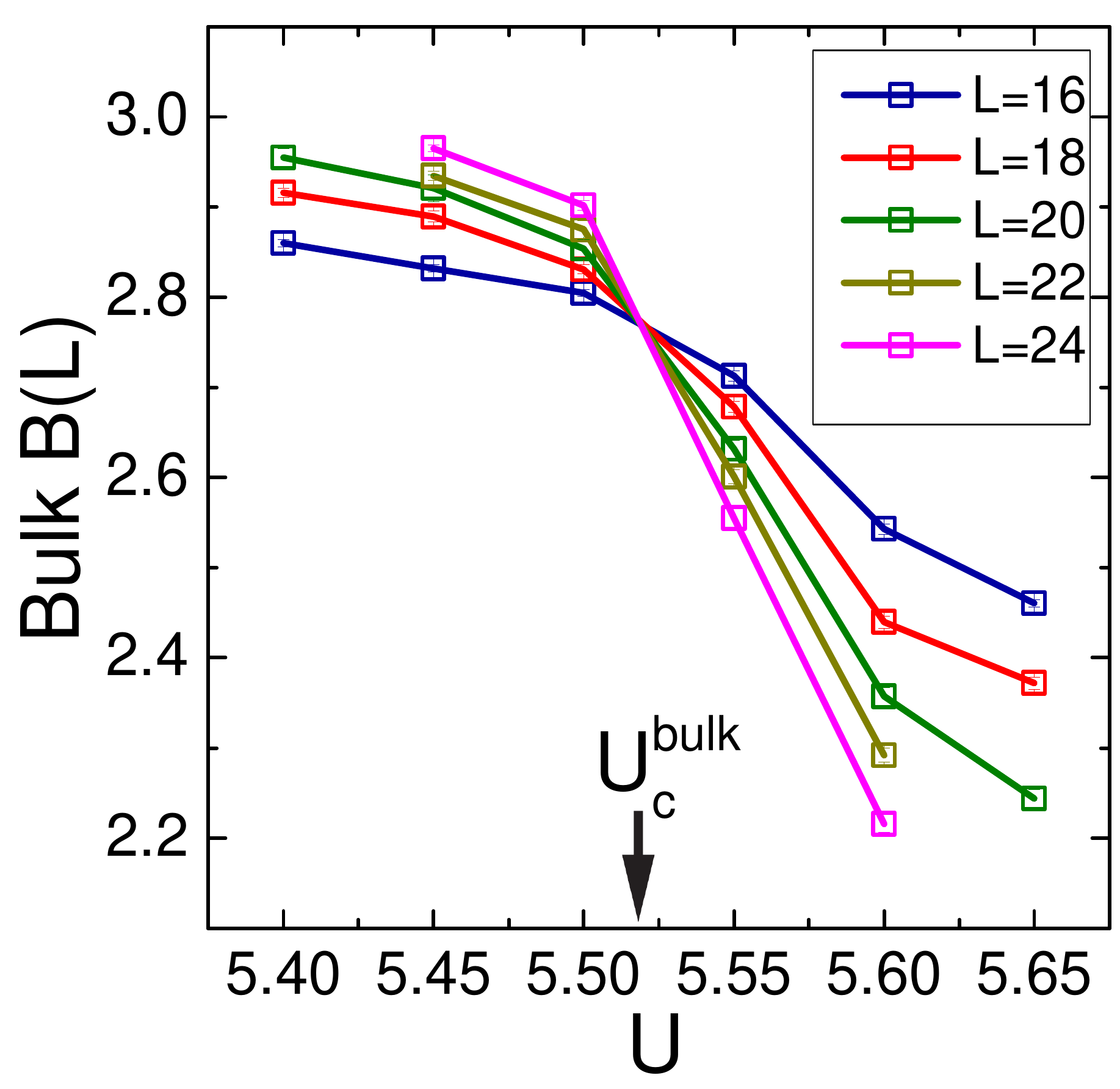}}
\subfigure[]{\includegraphics[height=3.0cm]{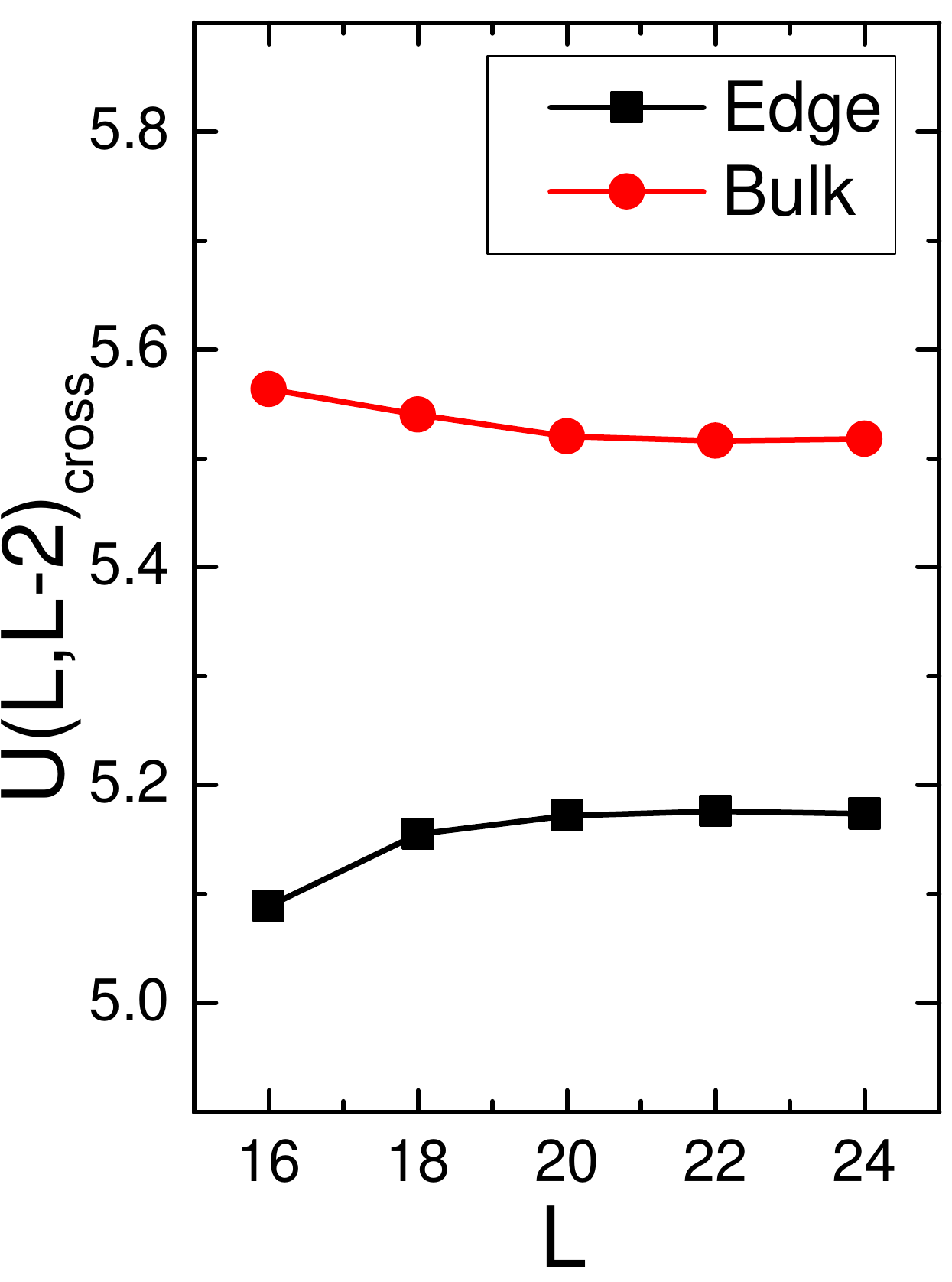}}					
\caption{QMC results clearly showing an EQCP of the interacting topological superconductor by tuning only a {\it single} parameter, the Hubbard interaction $U$. (a) For $\Delta$=0.4 and $\mu$=$-$0.5, the Binder ratio on the edge obtained from QMC shows that the edge QCP occurs $U^\textrm{edge}_c\approx 5.18$; (b) The Binder ratio in the bulk shows that the bulk QCP happens at a stronger interaction $U^\textrm{bulk}_c \approx 5.52$ with $U^\textrm{bulk}_c>U^\textrm{edge}_c$. (c) The plot of crossing points $U^{(L,L-2)}_\text{cross}$ between the two Binder ratios of system size $L$ and $L-2$, for $L=16,18,20,22,24$. It is clear that the edge QCP is separated from the bulk QCP.     }
\label{spinful}
\end{figure}

{\bf Edge quantum critical point:} It is expected that a strong attractive Hubbard interaction would generate a finite singlet pairing. Consequently, as the value of $U/t$ is increased, the TSC with only triplet pairing should encounter a quantum phase transition into a topologically trivial superconductor with a finite singlet-pairing component, namely $\Delta_s$=$\langle c^\dag_{i\A}c^\dag_{i\V}\rangle$$\neq$ 0, which spontaneously breaks the spin-parity symmetry $P$. Moreover, a pure imaginary $\Delta_s$, which spontaneously breaks the time-reversal symmetry $T$, could gain more condensation energy than real $\Delta_s$ \cite{Assaad-2016}. If an EQCP with spontaneous $T$ breaking occurs while the bulk remains gapped and topologically nontrivial, it was theoretically predicted that spacetime SUSY should emerge at the EQCP of the helical Majorana edge states \cite{Ashvin2014}.

So far, whether an EQCP in interacting topological phases including topological superconductors exists remains open. To investigate the nature of broken symmetry and the possibility of an EQCP in the interacting TSC, we perform the large-scale projector (namely, zero-temperature) QMC simulations \cite{Sugiyama-86,Sorella-89,White-89,Li2015} of the sign-problem-free model in \Eq{tsc} on the $L$$\times$$L$ square lattice with largest $L=24$. We evaluate the Binder ratio to determine the exact quantum critical point of edge and bulk time-reversal symmetry breaking. At a critical point, the value of the RG-invariant quantity Binder ratio should be independent of system sizes, such that the critical point can be identified as the crossing point of the Binder ratio for different system sizes. The finite-size effects on the crossing points of the Binder ratio for different system sizes are rather weak \cite{binder}. By computing the Binder ratio $B(L)$ of the singlet-pairing order parameter $\langle c^\dag_{i\uparrow} c^\dag_{i\downarrow}\rangle$ on the edge and in the bulk of the square lattice, we first obtain the critical values $U_{c}^\text{edge}$ and $U_{c}^\text{bulk}$ of spontaneous symmetry breaking occurring on the edge and bulk, respectively (see the Supplementary Materials for details). For $\Delta$=0.4 and $\mu$=$-$0.5, as shown in \Fig{spinful}, the edge spontaneous symmetry-breaking happens indeed before the bulk does such that there is a finite interaction range where the edge symmetry-breaking occurs while the bulk still preserves all the symmetries. Moreover, for $U$$>$$U^\text{edge}_{c}$ the singlet-pairing $\Delta_s$ obtained from QMC is purely imaginary, indicting that the $\mathrm{T}$ is spontaneously breaking. Another manifestation of $\mathrm{T}$ breaking is the appearance of magnetic ordering on the edge, as shown in \Fig{schematic} (see the SM for details).

For $U$$>$$U^\text{edge}_c$, the edge Majorana fermion is gapped and acquires a finite mass due to the $T$ breaking. Note that for $U^\text{edge}_c$$<$$U$$<$$U^\text{bulk}_c$, $T$ breaking appears only on the edge, but not in the bulk. At $U$=$U^\text{bulk}_{c}$, the BQCP with breaking time-reversal symmetry appears and the bulk quantum transition  is in the Ising university class in 2+1 dimensions, as shown in \Fig{schematic}. \\

\begin{figure}
\subfigure[]{\includegraphics[height=4.1cm]{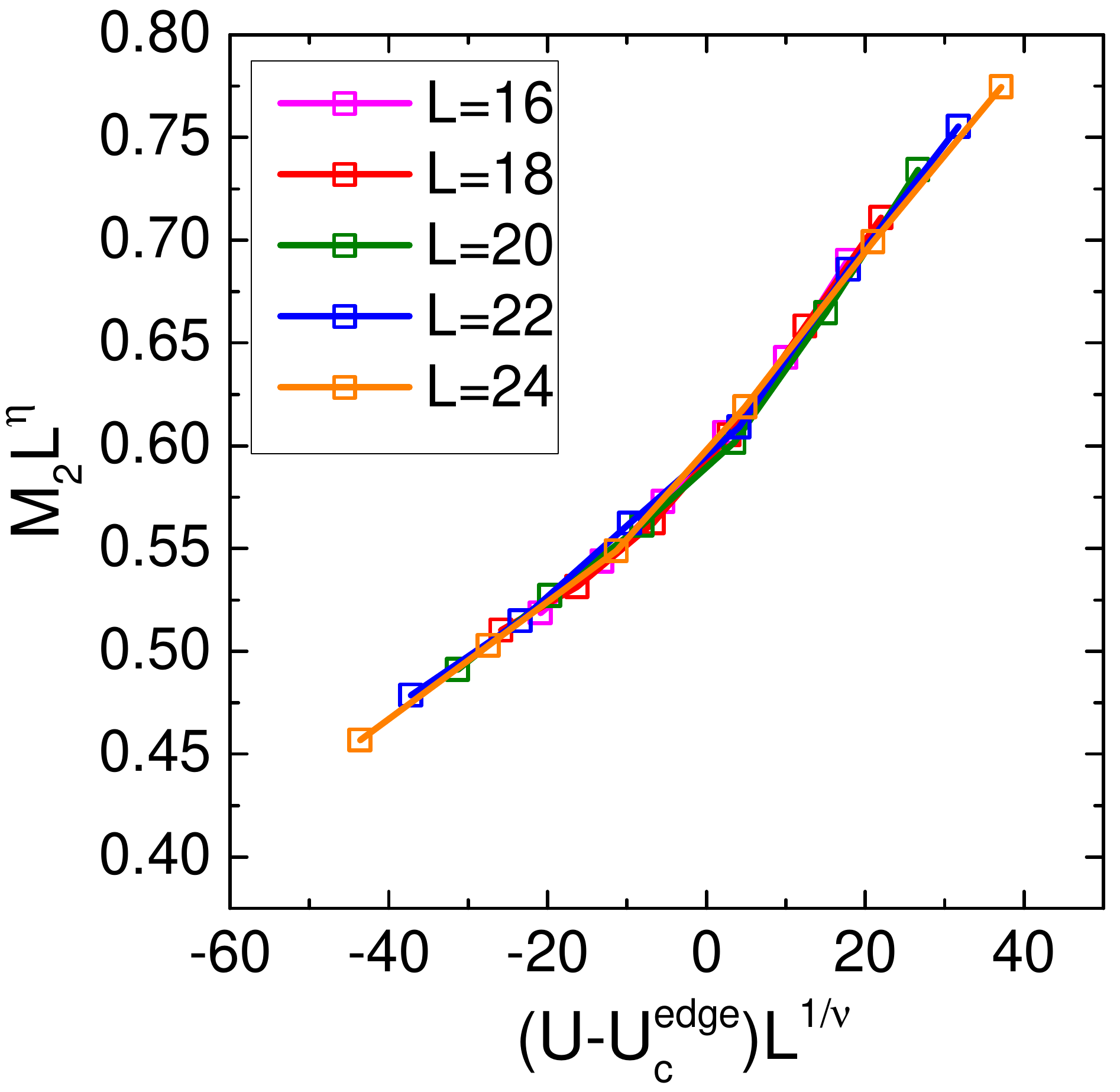}}~
\subfigure[]{\includegraphics[height=4.1cm]{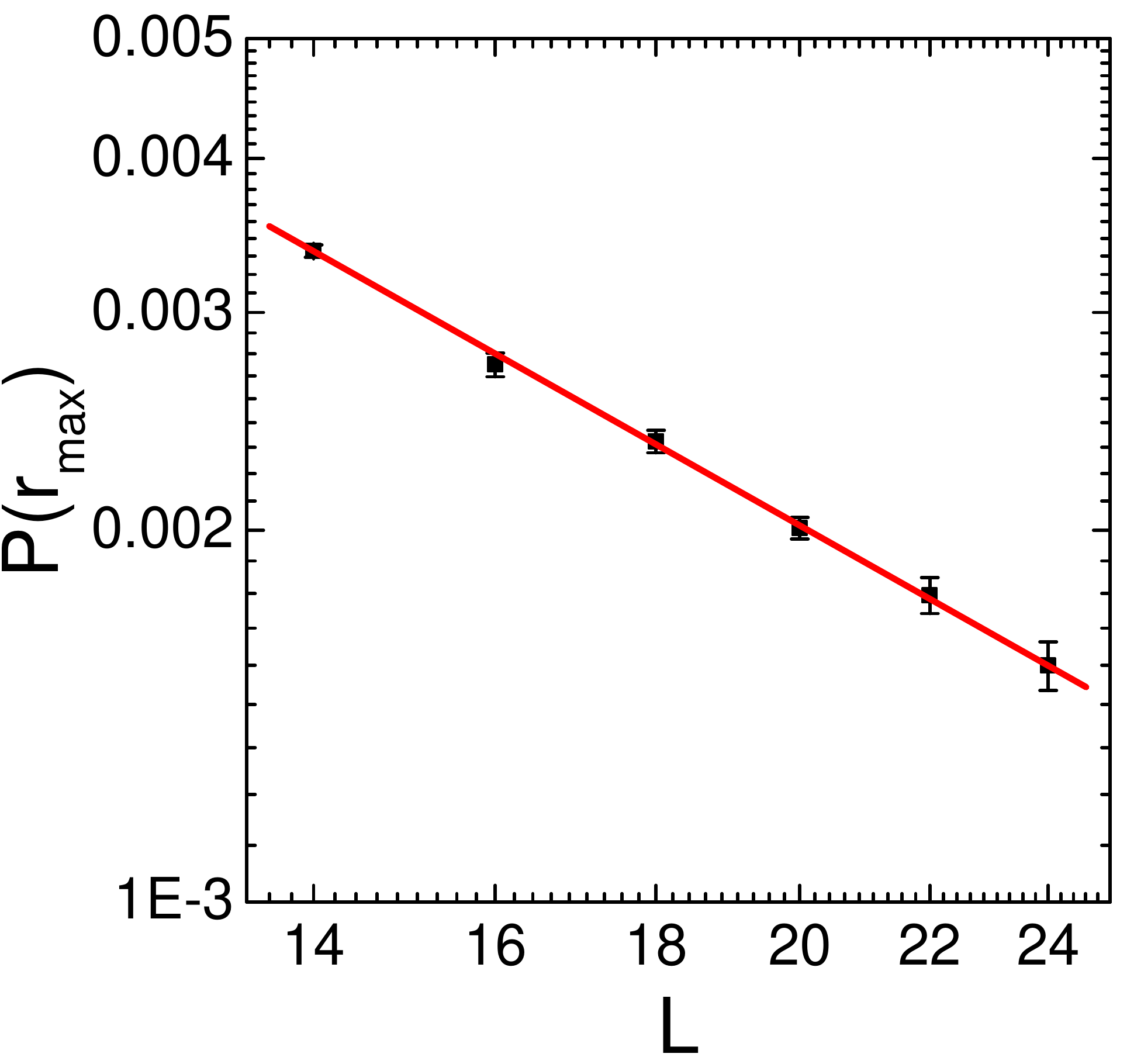}}	
\caption{QMC results of the critical exponents at the EQCP. (a) By employing data collapse to fit the scaling function of the structure factors near the EQCP $U=U^\textrm{edge}_c$ with $L=16,18,20,22,24$, we obtain $\eta_\textrm{QMC} = 0.43 \pm 0.03$ and $\nu_\textrm{QMC}=0.56\pm 0.03$, which is perfectly consistent with exact values of $\eta_\textrm{SUSY}=2/5$ and $\nu_\textrm{SUSY}=5/9$, convincingly indicating emergent SUSY at the EQCP. (b) Fermion correlation at largest separation $r_{max} = \frac{L}{2}$ is plotted versus linear system size $L$, for $L=16,18,20,22,24$. The fermion anomalous dimension $\eta_\psi = 0.45 \pm 0.07$ is obtained from the slope of linear scaling function in ln-ln plotting, which equals to anomalous dimension of boson within the error bar.}
\label{exponent}
\end{figure}

{\bf Emergent SUSY:}
The edge quantum phase transition of $T$ breaking discovered by our QMC simulations can be described by the following effective field theory,
\bea
S_\text{edge}=\int d\tau dx
\Big[\bar\psi(\sigma^0\pa_\tau+\sigma^zv_f\pa_x)\psi+\phi(\pa^2_\tau+v^2_b\pa^2_x)\phi\nn\\ +r\phi^2+u\phi^4+g\phi\psi\sigma^y\psi\Big],~~~
\eea
where $\psi=(\psi_\A,\psi_\V)^T$ describes edge Majorana fermions, $\phi$ represents the order-parameter boson, and $v_f$ ($v_b$) labels initial fermion (boson) velocity ($v_f$ and $v_b$, which are in general not equal). At the EQCP (namely, $r$=$0$), the RG analysis \cite{Ashvin2014} shows that there exists a fixed point where $v^\ast_f$=$v^\ast_b$=$v^\ast$ and $(g^\ast)^2$=$u^\ast$. The action at the fixed point is invariant under the following supersymmetric transformations: $\delta\phi=\bar\psi \theta$ and $\delta\psi= i \gamma^\mu\pa_\mu\phi \theta+g^\ast\phi^2\bar\theta$, where $\theta$ is a two-component spinor of real Grassmann variables that parametrizes the transformation and $\bar\psi=\psi^\dag\gamma^0$; in other words, spacetime SUSY emerges at the infrared limit from RG analysis. Nonetheless, so far it has not been tested from any unbiased simulations of two-dimensional microscopic models supporting the EQCP.

It is thus of urgent importance to verify the emergence of SUSY from the unbiased and numerically exact QMC simulations \cite{Melko2013}.  For the SUSY theory discussed above, the {\it exact} values of boson anomalous dimension $\eta_\phi$, fermion anomalous dimension $\eta_\psi$, and the correlation exponent $\nu$ are known theoretically: $\eta^{SUSY}_\phi$=$\eta^{SUSY}_\psi$=2/5 and $\nu^{SUSY}$=5/9 according to the superconformal symmetry \cite{Shenker-84}. To directly verify whether the emergent SUSY occurs in the 2D microscopic model we compute the critical exponents $\eta_\phi$, $\eta_\psi$, and $\nu$ at the EQCP by QMC simulations and compare them with the exact values of the putative 1+1 dimensional $\mathcal{N}$=1 SUSY or 1+1 dimensional tricritical Ising universality class \cite{Ashvin2014,Shenker-84}. Remarkably, as shown in \Fig{exponent}, our large-scale QMC simulations of the interacting TSC model with $\Delta$=$0.4$ and $\mu$=$-0.5$ give rise to the following critical exponents at the EQCP ($U$=$U^\text{edge}_c$): $\eta^\text{QMC}_\phi$=0.43$\pm$0.03, $\eta_\psi^\text{QMC}$=0.45$\pm$0.07, and $\nu^\text{QMC}$=0.56$\pm$0.03, all of which are equal to exact values obtained from the predicted emergent SUSY. (See Supplementary Material for details of deriving those critical exponents from finite-size scaling analysis.) Similar consistency is also obtained in QMC simulations of the TSC model with $\Delta$=0.3 \cite{SM}.
Consequently, our QMC simulations provide convincing evidence that the ${\cal N}$=1 SUSY indeed emerges generally at the EQCP of the two dimensional interacting TSC. We emphasize that the simulations here are directly studying a microscopic TSC model in two-dimensional with local time-reversal symmetry instead of an effectively one-dimensional model with nonlocal symmetry studied in \Ref{Ashvin2014}. \\

\begin{figure}[t]
\subfigure[]{\includegraphics[height=4.1cm]{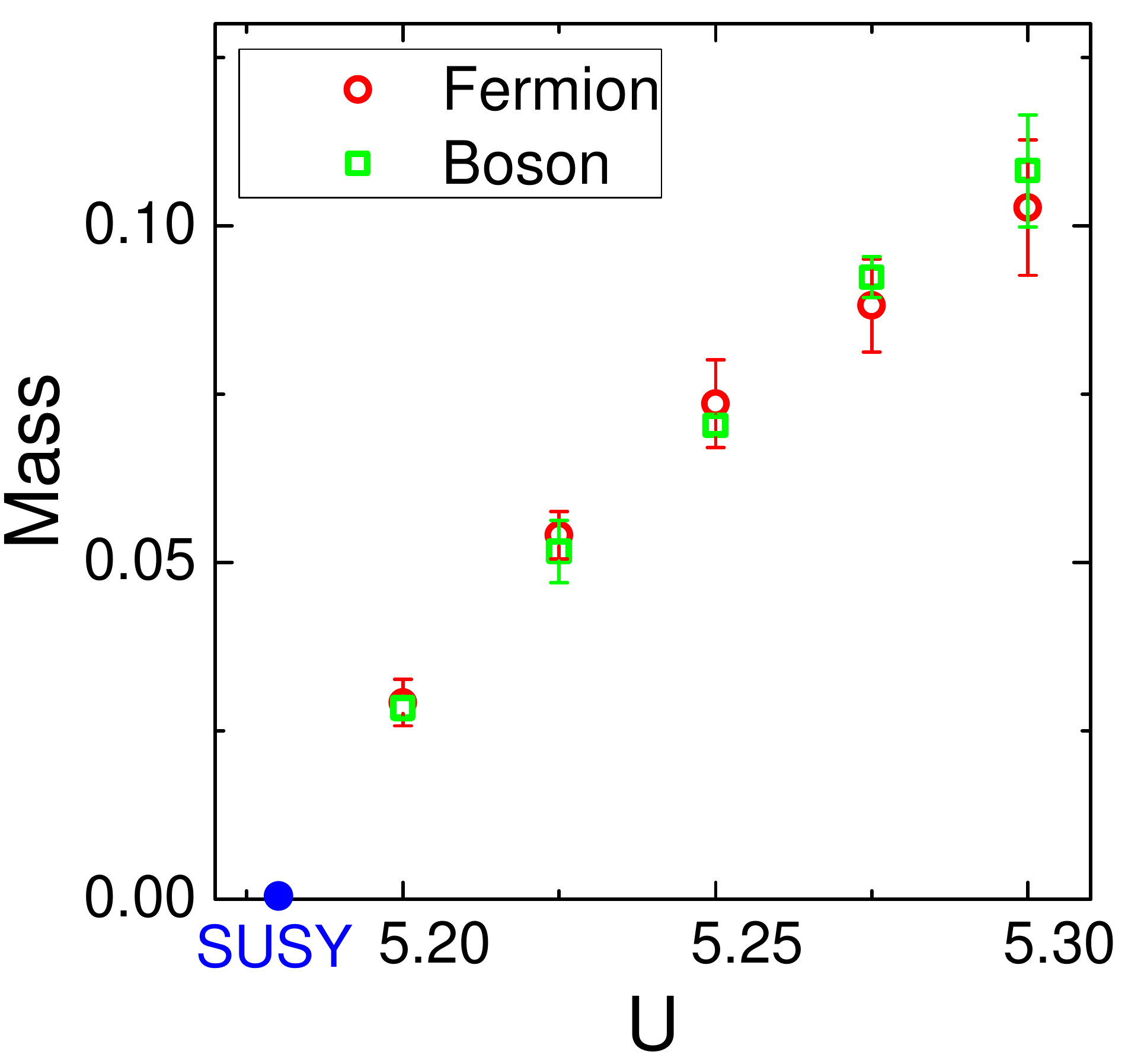} \label{massp3}}~
\subfigure[]{\includegraphics[height=4.1cm]{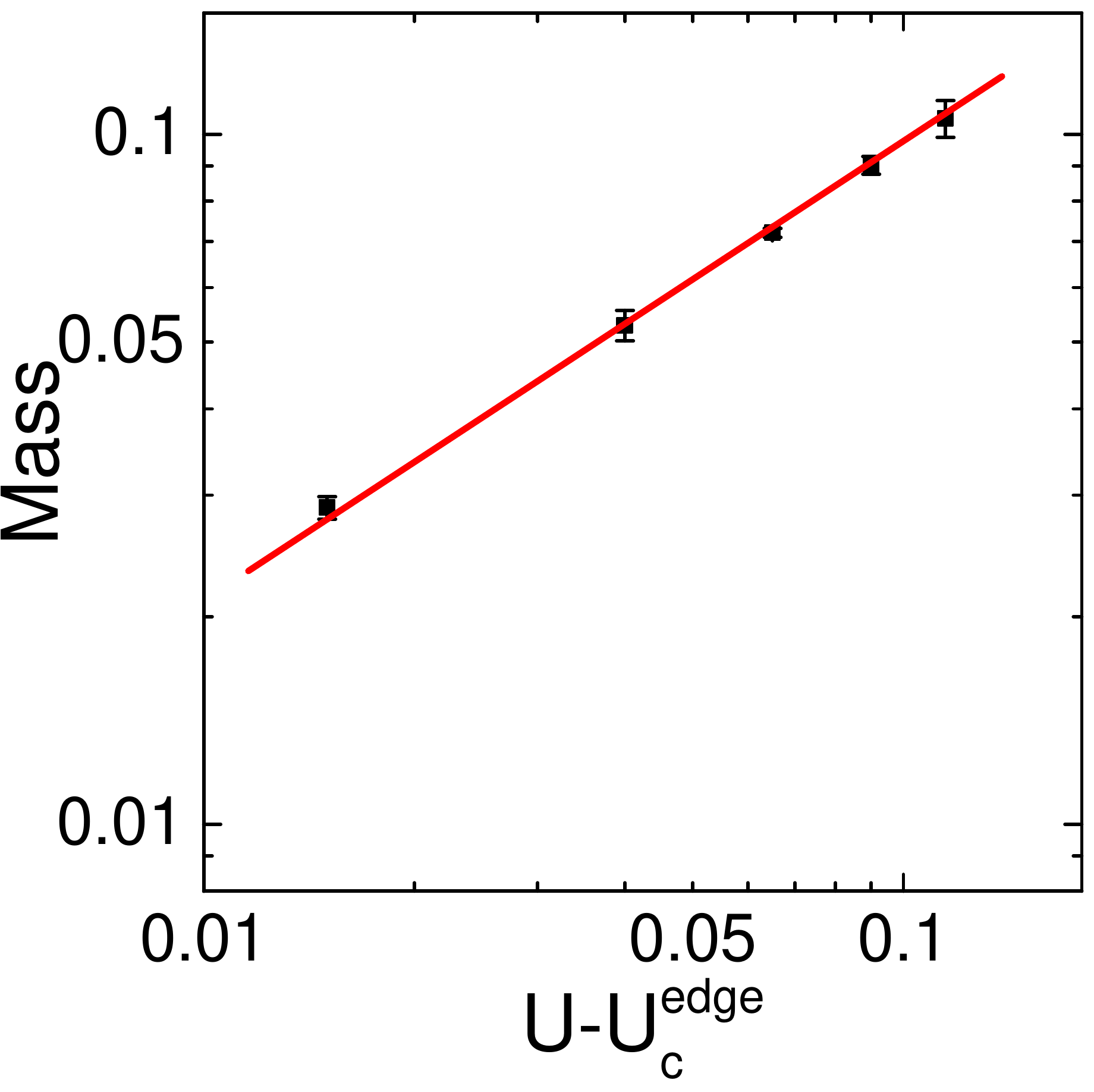} \label{massp4}}
\caption{QMC results showing emergent SUSY at the edge QCP of the interacting topological superconductor of spin-1/2 electrons. (a) Close to EQCP, fermion mass and boson mass equal to each other within error bar; (b) two masses can be fitted by the same scaling function $m_f=m_b\propto (U-U^\text{edge}_c)^{\nu} $ close to the EQCP.}
\label{mass}
\end{figure}

{\bf Equal mass of fermions and bosons:}
In a supersymmetric theory, particles and their superparticles share the same mass and other internal quantum numbers except spin. Consequently, emergent SUSY at the EQCP dictates that the edge Majorana fermion and its bosonic superpartner have equal masses at and near the EQCP. This hallmark phenomenon of emergent SUSY has not been observed in quantum materials or microscopic models so far. Fortunately, masses of fermions and bosons can be computed in QMC simulations. In this paper, from computing the following imaginary-time Green¡¯s function on the edge by QMC in the broken symmetry phase,
\bea
G^f_{k}(\tau)= \big\langle c^\dag_{k\sigma}(\tau) c_{k\sigma}(0) \big\rangle \propto \exp(-\tau m_f)+\cdots
\eea
where $k$ is the edge crystal momentum with $k=0$ and $\tau$ is sufficiently large, we are able to extract the mass $m_f$ (or equivalently gap) of edge Majorana fermions generated by $T$ breaking. Similarly, the boson's mass can also be computed from evaluating the edge boson Green's function $G^b_k(\tau)$=$\langle\hat \phi^\dag_k(\tau)\hat\phi_k(0)\rangle$$\propto$$\exp(-\tau m_b)+\cdots$,  where $\phi_k$ is the Fourier transform of the edge order parameter $\hat \phi_j$=$i(c^\dag_{j\A}c^\dag_{j\V}-c_{j\V}c_{j\A})$, in the $T$ breaking phase \cite{SM}. The values of these two masses in the thermodynamic limit are obtained by finite-size scaling, as shown in the Supplemental Material. The error bars of the values in \Fig{massp3} represent the standard errors of fitting (see \cite{SM} for details). Sufficiently close to the EQCP, we find that the two masses equal to each other within numerical error bar, namely $m_f$=$m_b$, as shown in \Fig{mass}(a).  The observation that fermions and bosons have equal dynamically generated masses provides a further evidence of emergent SUSY at the EQCP. When the system is sufficiently away from the EQCP, the two masses differ apparently indicating the breaking of SUSY, which could shed some light on understanding possible SUSY breaking in nature.

\begin{figure}[t]
\includegraphics[width=8.cm]{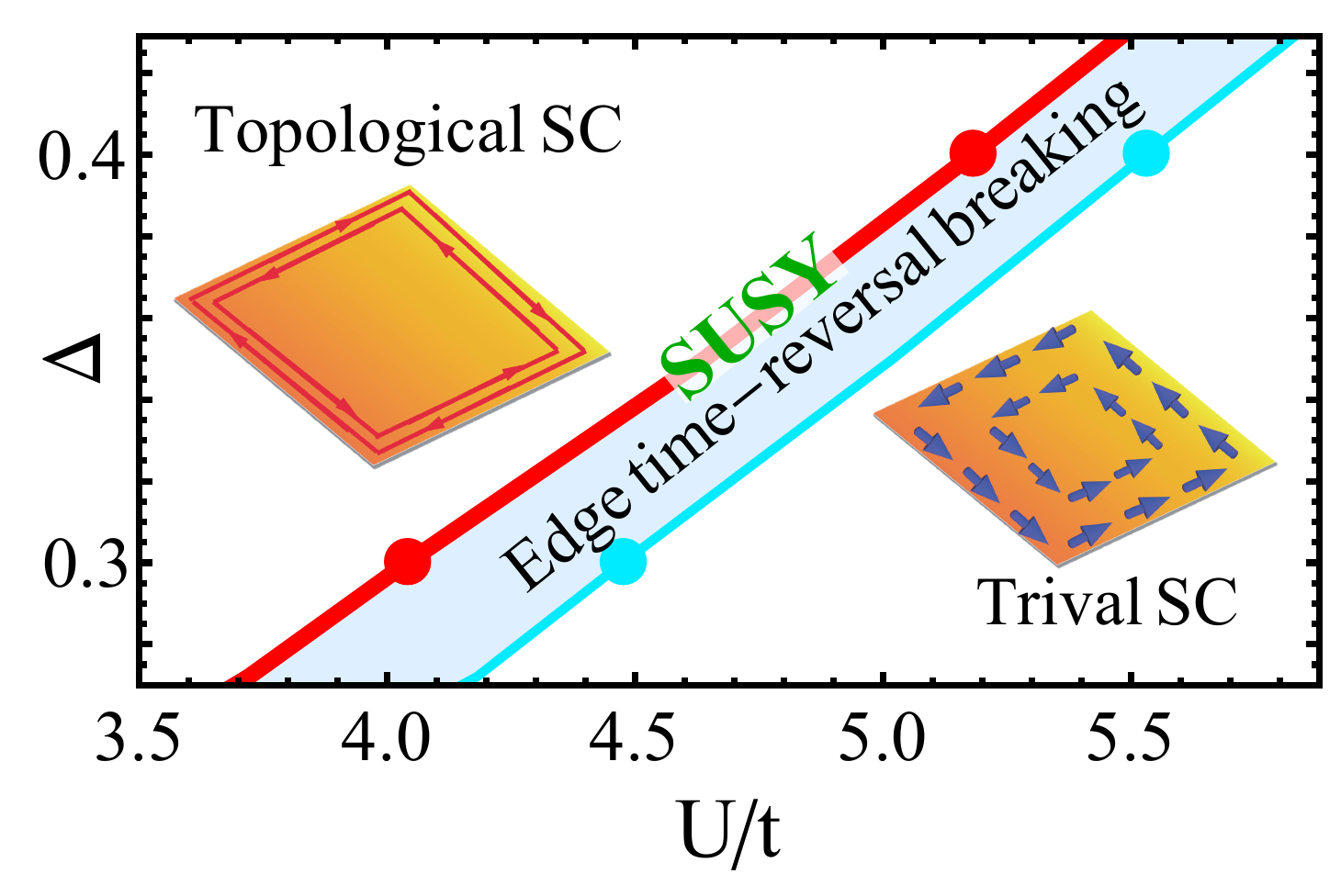}
\caption{The quantum phase diagram of interacting topological superconductors. Here the dots are obtained by QMC simulations. The EQCP (red line) and BQCP are generically separated for any finite $\Delta$. The spacetime $\mathcal{N}$=1 SUSY emerges at the EQCP. }
\label{diagram}
\end{figure}

Exactly at the EQCP, both $m_f$ and $m_b$ certainly vanish. Close to the EQCP with the emergent SUSY, the scaling of masses is given by
\bea\label{massscaling}
m_f=m_b\propto (U-U^\text{edge}_c)^{\nu},
\eea
where $\nu$=$\nu^\text{SUSY}$=$5/9$ owing to the SUSY. The masses obtained from the QMC simulations are reasonably consistent with this scaling, as shown in \Fig{mass}(b), providing further support to the emergent SUSY at the EQCP of the interacting two-dimensional topological superconductors, whose global phase diagram as a function of triplet pairing $\Delta$ and on-site interaction $U$ is shown in \Fig{diagram}.\\

{\bf Concluding remarks:}
One largely open but important issue is how to possibly realize emergent SUSY in condensed matter or cold atom systems. As there are already promising proposals to realize time-reversal-invariant TSC in two dimensions \cite{dhlee2012,kane2013,nagaosa2013,hyao2015}, tuning a single parameter such as the Hubbard interaction studied in this work should suffice to achieve the EQCP and emergent SUSY. Another promising way is to employ ultracold atoms loaded into an optical lattice, where relatively strong on -site Hubbard attractions can be achieved by tuning the system close to Fechbach resonance \cite{coldreview}.

The edge quantum phase transition with time-reversal breaking magnetic orders may be detected by measuring local magnetic fields via experimental probes such as scanning superconducting quantum interference devices and scanning magnetic force microscopy or by measuring the gap of edge Majorana fermions by scanning superconducting quantum interference devices (STM). The bulk symmetry breaking may be detected by polar Kerr rotation \cite{kapitulnik2014,kivelson2016}. Around the EQCP, the critical exponent $\nu$ can be inferred from the edge Majorana fermion gap measured by STM according to \Eq{massscaling}, which can be further compared to the exact value $5/9$ predicted by SUSY. Moreover, exactly at or sufficiently close to the EQCP, another experimental signature of the emergent SUSY is the local density of state(DOS) at the EQCP that can be measured by STM: $\rho(\omega)$$\propto$$|\omega|^{2/5}$, which is qualitatively different from the constant DOS ($\propto$$|\omega|^0$) of noninteracting edge Majorana fermions. Another interesting consequence of the edge symmetry breaking of the two-dimensional TSC is the emergence of a Majorana zero mode \cite{beenakker-13} localized at the domain boundary of the edge magnetic orders, which may be detected by STM.

In conclusion, from numerically exact QMC simulations, we have shown that an EQCP can be achieved in a minimal microscopic model of two-dimensional interacting TSC by tuning only a single parameter. An EQCP has been convincingly established in two-dimensional lattice models, which could shed light on studies of exotic quantum critical points \cite{Senthil2004,Sandvik2016,li2015fiqcp,dhlee2011,herbut2016,YuanyaoHe2016, assaad2011, wu2011, cenke2015}. Moreover, our unbiased simulations show convincing evidence of emergent SUSY at the EQCP by both obtaining consistent critical exponents and demonstrating equal dynamically generated masses of fermions and bosons. We believe that the results presented here can not only lend concrete support to potentially realize emergent spacetime SUSY in quantum materials but also shed light on the intriguing interplay among correlation, topology, and symmetry.

{\it Acknowledgements}: We sincerely thank Shou-Cheng Zhang and Shao-Kai Jian for helpful discussions. This work was supported in part by the National Natural Science Foundation of China under Grant No. 11474175 (Z.-X. L., Y.-F. J., and H. Y.), by the Ministry of Science and Technology of China under
Grant No. 2016YFA0301001 (H. Y.) and by the National Thousand-Young-Talents Program (H. Y.).

\begin{widetext}
\section{supplemental materials}

\renewcommand{\theequation}{S\arabic{equation}}
\setcounter{equation}{0}
\renewcommand{\thefigure}{S\arabic{figure}}
\setcounter{figure}{0}
\renewcommand{\thetable}{S\arabic{table}}
\setcounter{table}{0}

\subsection{I. Details of the projector Majorana QMC simulations}
We use projector QMC in Majorana representation to investigate the interacting topological superconductors described by the Hamiltonian in Eq. (1). In projector QMC, the expectation value of an observable $O$ in the ground state can be evaluated as:
\bea
\frac{ \bra{\psi_{0}} O \ket{\psi_0}}{ \avg{\psi_{0} \mid \psi_{0} } }   = \lim_{\theta\rightarrow \infty} \frac{ \bra{\psi_T}e^{-\theta H } O e^{-\theta H} \ket{\psi_T}}{ \bra{\psi_T} e^{-2\theta H} \ket{\psi_T}},
\eea
where $ \psi_0 $ is the true ground state wave function and $ \psi_T$ is a trial wave function which should have a finite overlap with the true ground state wave function. In our QMC simulations, the imaginary-time projection parameter is $\theta = 60/t$ for the systems with periodic boundary conditions. In the cases of cylinder boundary conditions, most systems are computed using $\theta = 100/t$ and some systems with large systems size or near critical points are computed using $\theta = 120/t$. We have checked that results stay nearly the same for larger $\theta$ which ensures desired convergence. We set $\Delta \tau =0.05$ and the results do not change if we use smaller $\Delta \tau$. Because of the absence of sign-problem, we can perform large-scale QMC simulations. In the computation of the bulk quantum phase transition, we use periodic boundary condition. In the computation of edge phase transition, we use periodic boundary condition in $x$ direction and open boundary condition in $y$ direction. The largest system size has $N=L^2$ sites with $L=24$.

\subsection{II. The Binder ratio and finite size scaling analysis for the bulk and edge QCP}

\begin{figure}[b]
\subfigure[]{\includegraphics[height=3.9cm]{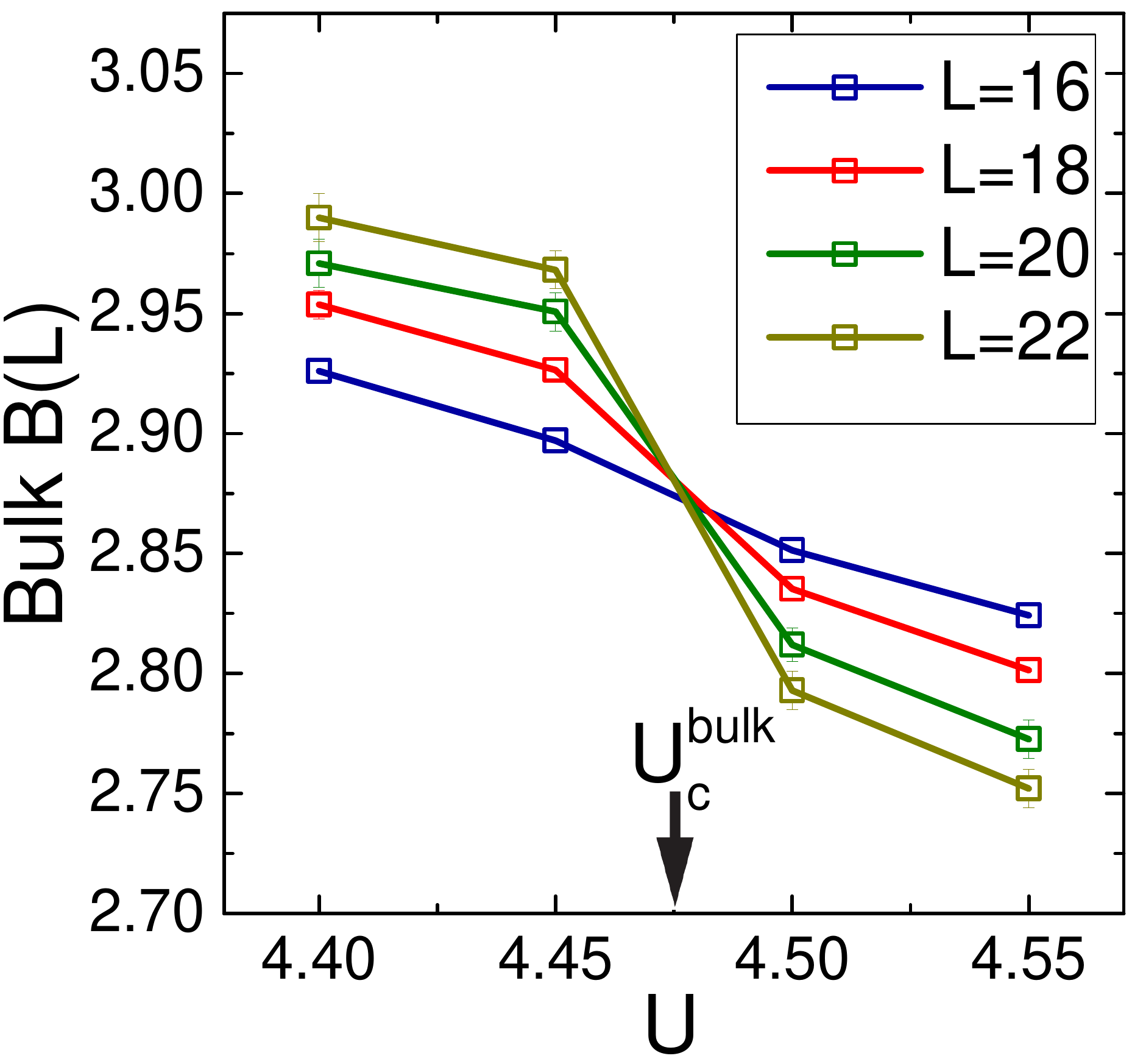}}
\subfigure[]{\includegraphics[height=3.9cm]{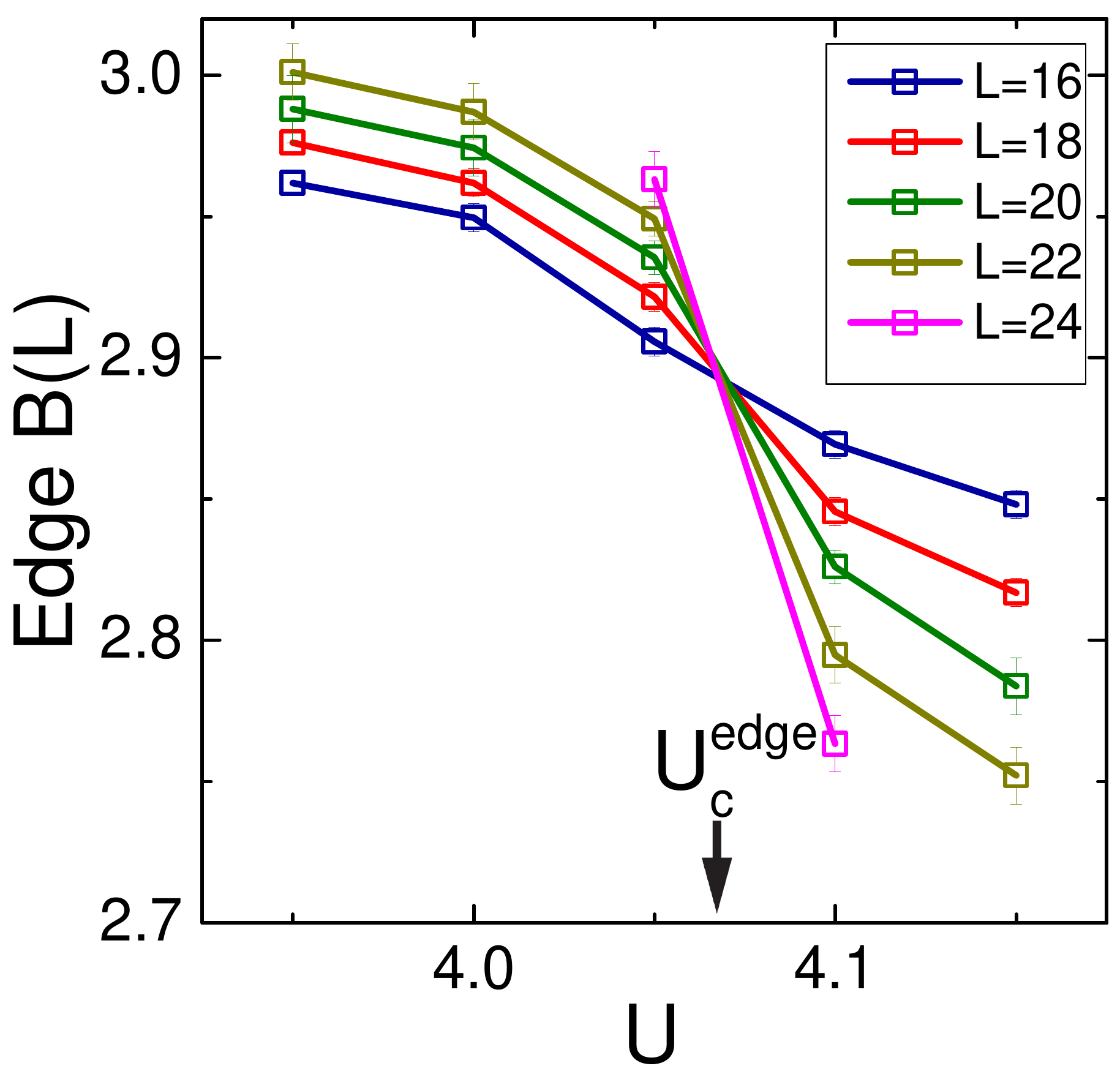}}\\
\subfigure[]{\includegraphics[height=3.9cm]{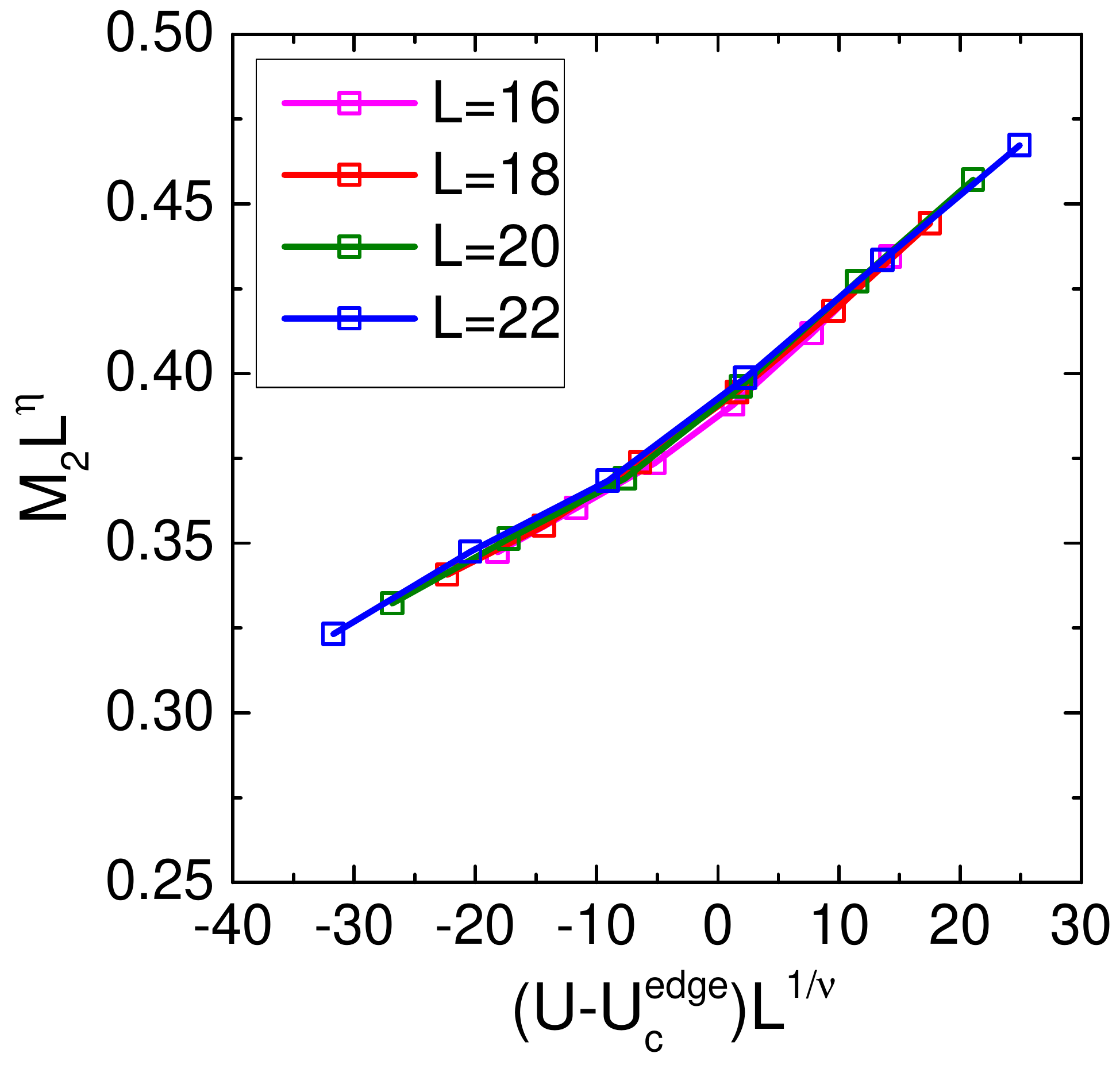}}
\subfigure[]{\includegraphics[height=3.9cm]{fermioncorrelation_p3}}
\caption{QMC results of Binder ratio and critical exponents at the EQCP for $\Delta = 0.3$ and $\mu=-0.5$. (a) Bulk Binder ratio shows that BQCP occurs at $U_c^\text{bulk} \approx 4.48 $; (b) Edge Binder ratio shows that EQCP occrus at $U_c^\text{edge} \approx 4.07$ with $U_c^\text{edge}<U_c^\text{bulk}$; (c) By employing the data collapse analysis to fit the scaling function of the structure factor near EQCP, $\eta_{\phi} = 0.42 \pm 0.03  $ and $\nu =  0.57 \pm 0.04 $ are obtained; (d) From the scaling of fermion correlation function on edge, we obtain the fermion anomalous dimension $\eta_{\psi} = 0.46 \pm 0.08$).}
\label{p3}
\end{figure}

As a powerful numerical technique, the Binder ratio $B(L) = \frac{M_4}{M_2^2}$ is frequently used to study quantum phase transitions, where $M_2$ and $M_4$ are the second-order and fourth-order moments associated with the order parameter describing the quantum phase transition in question. For the quantum phase transition breaking time-reversal symmetry, the order parameter is pure-imaginary singlet-pairing $\langle i c^\dagger_{i\uparrow}c^\dagger_{i\downarrow} - i c_{i\downarrow}c_{i\uparrow}\rangle$. For this quantum phase transition, the $M_2$ is the pure-imaginary singlet-pairing structure factor: $M_2 = \sum_{ij} \frac{1}{N^2} \big\langle (i c^\dagger_{i\uparrow}c^\dagger_{i\downarrow} - i c_{i\downarrow}c_{i\uparrow})(i c^\dagger_{j\uparrow}c^\dagger_{j\downarrow} - i c_{j\downarrow}c_{j\uparrow})\big\rangle $ and $M_4 = \sum_{ijkl} \frac{1}{N^4} \big\langle (i c^\dagger_{i\uparrow}c^\dagger_{i\downarrow} - i c_{i\downarrow}c_{i\uparrow})(i c^\dagger_{j\uparrow}c^\dagger_{j\downarrow} - i c_{j\downarrow}c_{j\uparrow}) (i c^\dagger_{k\uparrow}c^\dagger_{k\downarrow} - i c_{k\downarrow}c_{k\uparrow})(i c^\dagger_{l\uparrow}c^\dagger_{l\downarrow} - i c_{l\downarrow}c_{l\uparrow}) $. In the computation of phase transition at bulk, we simulate $M_2$ and $M_4$ by a summation over all the sites. While in the computation of phase transition at edge, we only sum over the sites at the boundary ($y=1$ and $y=L_y$). In disordered phase, the Binder ratio increases as the systems size in increased, while in ordered phase its trend is opposite. The Binder ratio for different $L$ should cross at the critical point.

To obtain the critical exponents of the edge and bulk quantum phase transitions, we perform finite-size scaling analysis. Close to quantum critical points, the structure factor $M_2$ of the time-reversal breaking singlet-pairing satisfies the following scaling function:
\bea\label{scalingM2}
M_2 = L^{-d-\eta+1} {\cal F}( L^{\frac{1}{\nu}}(U-U_c)),
\eea
where we have implicitly assumed the dynamical critical exponent $z = 1$ for the quantum phase transition of the current model. Here, for the bulk phase transition, $d=2$; for edge phase transition, $d=1$. When sufficiently close to the quantum critical point, the structure factor $M_2$ should be collapsed to a single smooth scaling function, as in \Eq{scalingM2}, for different $L$ and $U$ by choosing the appropriate values of $\eta$ and $\nu$.

In the main text, we have shown the results for $\Delta$=0.4, for which the edge time-reversal symmetry breaking occurs at the interaction strength smaller than that of the bulk transition. Moreover, at the EQCP, the observed critical exponents and equal masses between fermions and bosons provide strong evidence that the EQCP features an emergent SUSY. The EQCP with emergent SUSY should be generic for the interacting topological superconductors. To confirm this, we further studied the case of $\Delta = 0.3$, for which we performed the same computations as $\Delta=0.4$ whose results have been shown in the main text. Indeed, we obtain similar results between $\Delta=0.3$ and $\Delta=0.4$, which leads to the conclusion that the EQCP in the interacting TSC model respects the same emergent SUSY. The followings are the details of the analysis. From Binder ratio analysis, it clearly shows that the edge QCP occurs at $U^\text{edge}_c = 4.09$ and the bulk QCP at $U_c^\text{bulk} = 4.48$, as shown in \Fig{p3}(a) and \Fig{p3}(b), respectively. There is a finite range of interactions where the edge spontaneously breaks $\mathrm{T}$ while the bulk preserves all the symmetries. From data collapse of $M^\text{edge}_2$, as shown in \Fig{p3}(c), we obtain the values of critical exponents at the EQCP: boson anomalous dimension $\eta_\phi = 0.42 \pm 0.03 $, fermion anomalous dimension $\eta_\psi = 0.46 \pm 0.08$, and $\nu = 0.57 \pm 0.04$.  These results are consistent with the exact values of the corresponding exponents of the SUSY theory.

\begin{figure}[b]
\subfigure[]{\includegraphics[height=4.1cm]{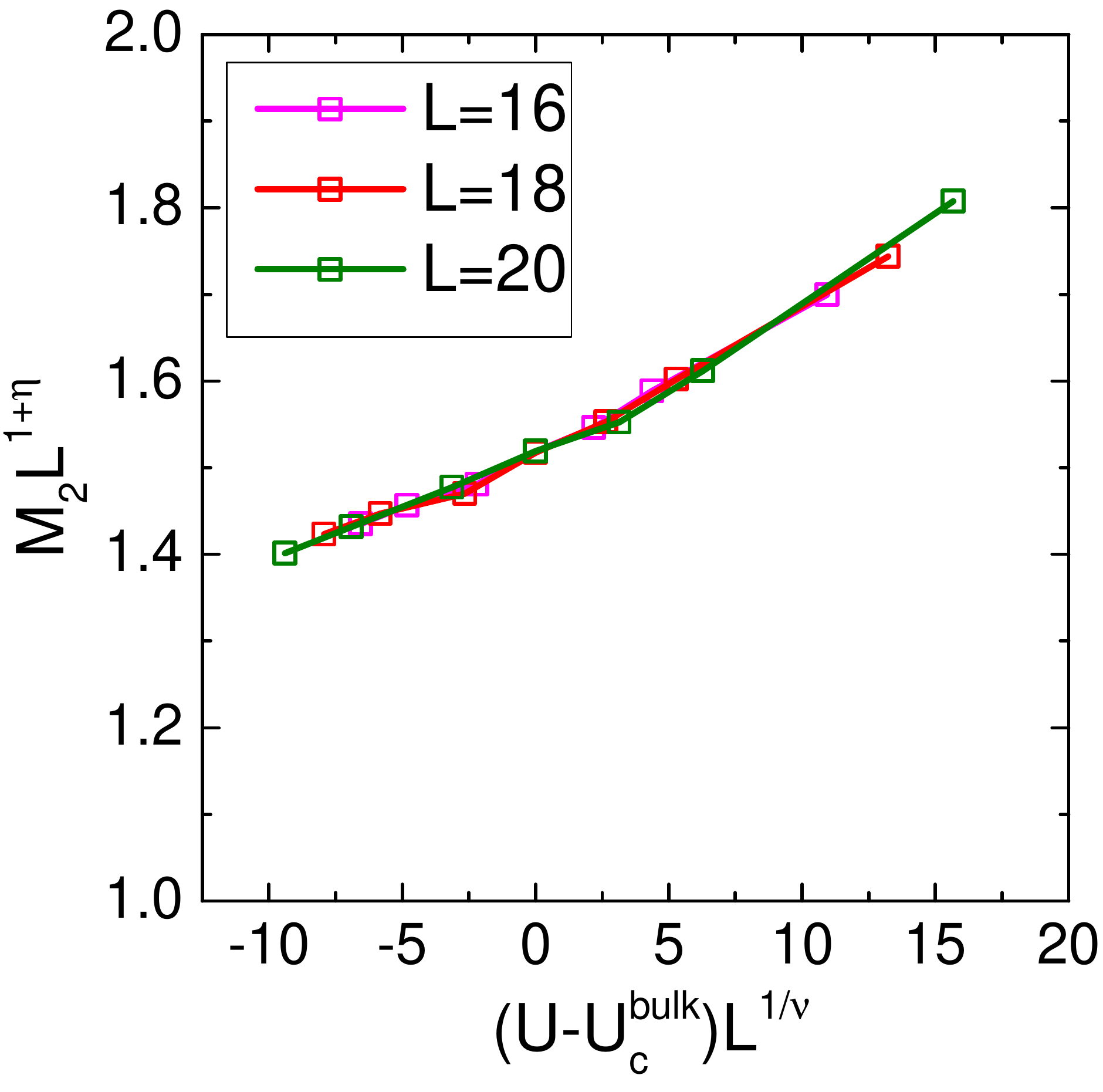}}
\subfigure[]{\includegraphics[height=4.1cm]{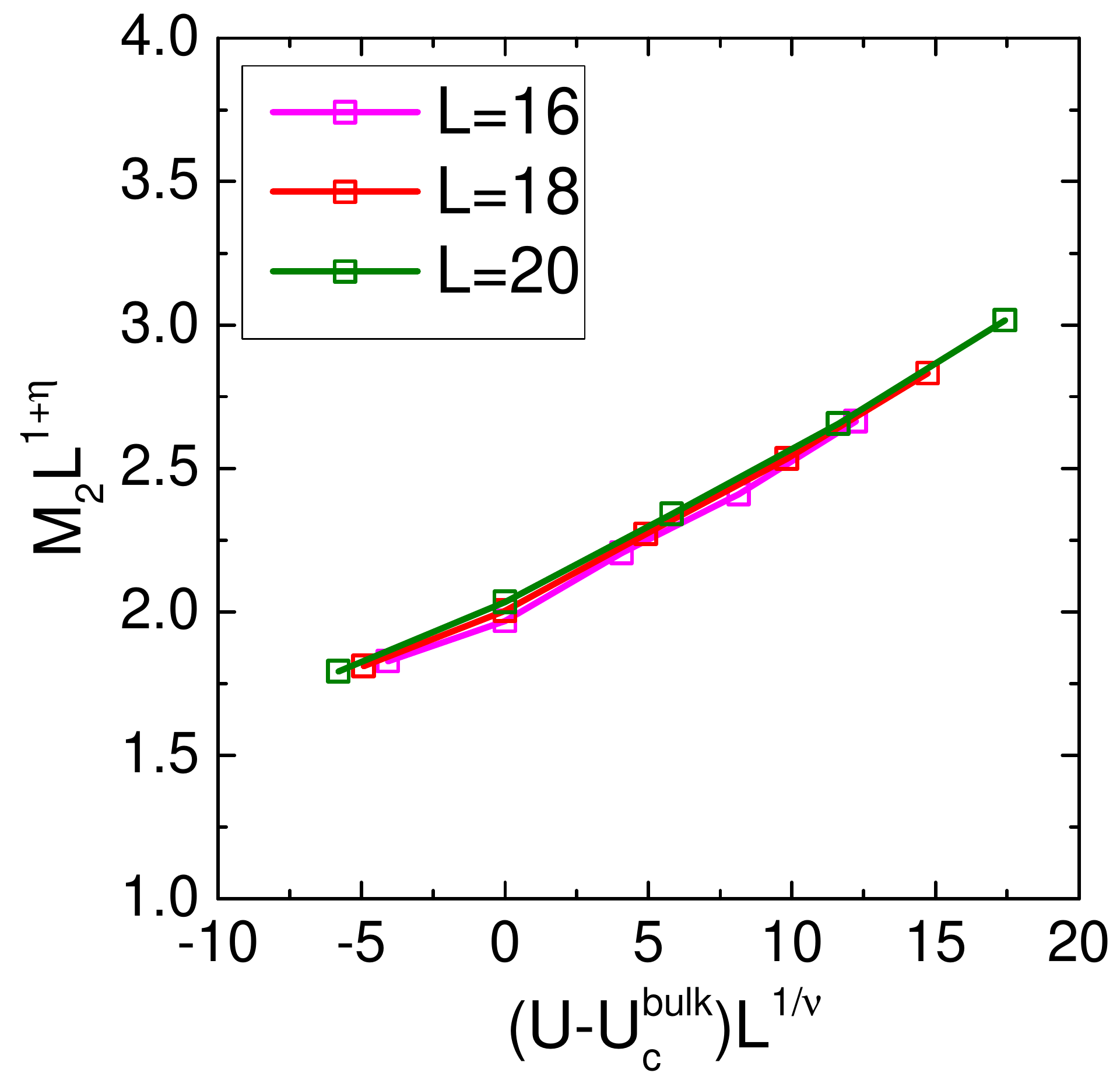}}
\caption{The finite-size scaling analysis at the BQCP, which belongs to the 3D Ising university class. (a) For $\Delta = 0.3$, the data collapse analysis of bulk $M_2$ gives rise to the following critical exponents of the BQCP: $\eta = 0.04 \pm 0.03 $, $\nu = 0.62 \pm 0.02 $. (b) Similar analysis of BQCP for $\Delta = 0.4$ gives rise to the critical exponents: $\eta= 0.02 \pm 0.03 $, $\nu = 0.63 \pm 0.03 $. For both cases, the critical behaviors are consistent with the 3D Ising university class.}
\label{bulkcritical}
\end{figure}

\subsection{III. Fermion and boson masses in the edge with time-reversal symmetry breaking}

When the time-reversal symmetry on the edges is broken, both fermions and bosons on the edges are gapped or massive. In the projector QMC simulations, the gap may be obtained through measuring the tails of the (imaginary) time displaced Green's function. To compute the gap of edge fermions, we measure single-particle Green's function:
\bea
G^f_k(\tau) = \sum_{\sigma=\uparrow,\downarrow}\avg{c^\dagger_{k\sigma}(\tau)c_{k\sigma}(0)},
\eea
where $c^\dagger_{k\sigma}(\tau) = e^{\tau H} c_{k\sigma}^\dagger e^{-\tau H}$ and $k$ is the edge crystal momentum. We use periodic boundary condition in $x$ direction and open boundary condition in $y$ direction. Here $k$ is the momentum in $x$ direction and $c^\dagger_k = \frac{1}{L_x}\sum_{x=1}^{L_x} c^\dagger_{(x,y=L_y)} e^{ik x}$.  The fermionic mass $m_{f}$ corresponds to the single-particle excitation energy at $k=0$, which can be obtained from $G^f_{k=0}(\tau) \propto e^{-\tau m_f}$ when imaginary-time $\tau$ is large enough. Similarly, the bosonic mass can also be obtained through imaginary-time displaced pair-pair correlation:
\bea
G^b_k(\tau) = \avg{\hat \phi^\dag_k(\tau)\hat\phi_k(0)},
\eea
where $\hat \phi^\dagger_k(\tau) = \frac{1}{L_x} \sum_{x=1}^{L_x} i\big[ c^\dagger_{(x,y=L_y)\uparrow}(\tau)c^\dagger_{(x,y=L_y)\downarrow}(\tau) -  c_{(x,y=L_y)\downarrow}(\tau)c_{(x,y=L_y)\uparrow}(\tau)\big] e^{i k x}$. The mass of bosons can be obtained from $G^b_{k=0}(\tau) \propto e^{-\tau m_{b}}$ when imaginary-time $\tau$ is large enough. 
We compute fermion and boson masses in systems with linear size $L = 10,12,14,16$ by fitting the tail of imaginary-time displaced single-particle Green's function and pair-pair correlation, respectively. Then, the values of masses in the thermodynamic limit are obtained by finite-size scaling using a linear function of $1/L$. We employ the standard method of least square fitting to extract the parameters in the linear function. The finite-size scaling of boson and fermion masses are presented in \Fig{mass_scaling}. The intercepts of the fitted linear function are values of boson and fermion masses in the thermodynamic limit. The error bars of masses are standard errors of the fitted intercepts.

\begin{figure}[t]
\subfigure[]{\includegraphics[height=3.22cm]{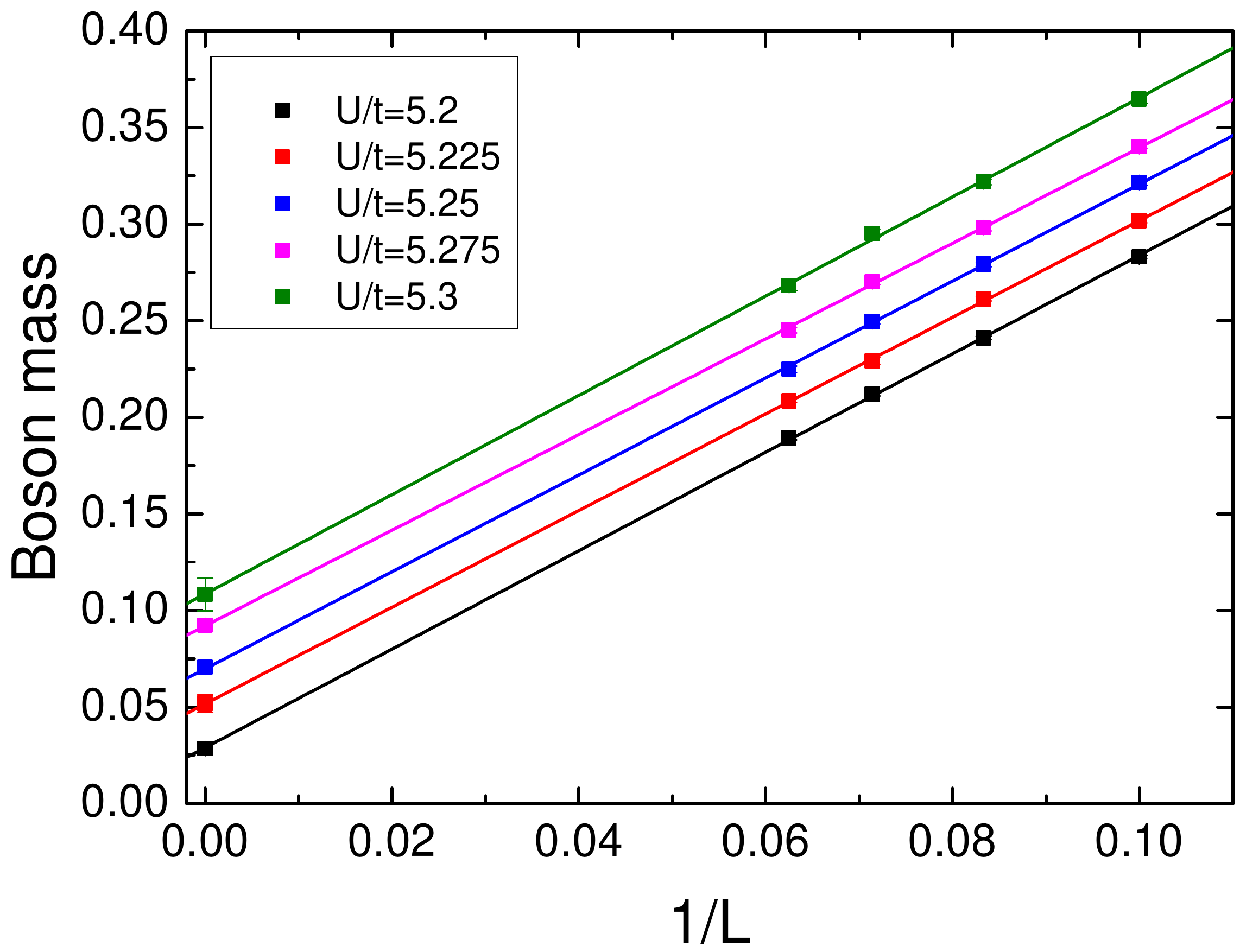}}
\subfigure[]{\includegraphics[height=3.22cm]{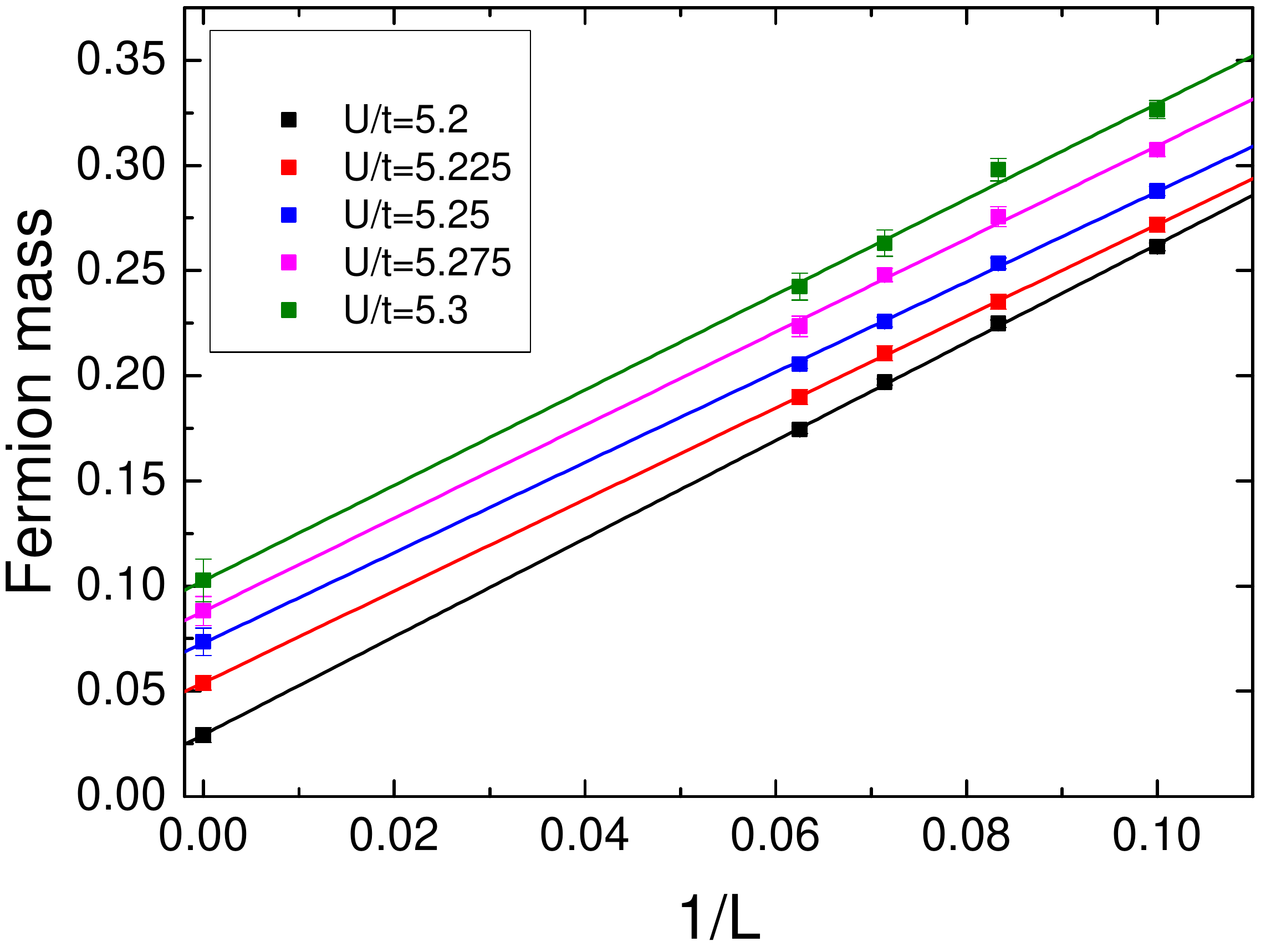}}
\caption{QMC results of the boson and fermion masses of the edge magnetic ordered phase close to the EQCP for $\Delta=0.4$. (a) The finite-size scaling of boson masses versus $1/L$ for $L=10,12,14,16$. A linear function is used in the fitting. The intercepts of linear functions are the values of boson masses in the thermodynamic limit. The error bars are standard errors of fitted parameters. (b) The finite-size scaling of fermion masses versus $1/L$ for $L=10,12,14,16$. The intercepts of linear functions are the values of fermion masses in the thermodynamic limit. }
\label{mass_scaling}
\end{figure}

\subsection{IV. Magnetic order in $\mathrm{T}$ breaking phase}

Besides the pure-imaginary singlet-pairing, $\mathrm{T}$ breaking can also manifest explicitly magnetic ordering. The magnetic order in the $\mathrm{T}$ breaking phase arises as the secondary order since it can be derived from the pure-imaginary singlet-pairing and helical $p$-wave triplet-pairing.  When we compute magnetic order on the edge, we use periodic boundary condition along the $x$-direction and open boundary condition along the $y$-direction. In the $\mathrm{T}$ breaking phase, magnetic moments on the edge are ordered in the $x$-direction due to the reflection symmetry on the $xz$-plane $M_y$ which map $y$ to $-y$. We compute the structure factor of magnetic order on the edge to further verify the $\mathrm{T}$ breaking on the edge: $S^\text{edge}(L) = \frac{1}{L^2} \sum_{ij}\langle  c^\dag_i\sigma^x c_i c^\dag_j\sigma^x c_j\rangle $, where $i,j$ are restricted to one edge. The finite-scaling scaling of the edge structure factor with $1/L$ shows that long-ranged magnetic order indeed appears when $U>U^\text{edge}_c$ (see \Fig{magnetic}(a)), which serves as another evidence of $\mathrm{T}$breaking on the edges.

Similarly, $\mathrm{T}$ breaking in the bulk can also be reflected by certain type of magnetic order. In our computation of bulk magnetic order, when periodic boundary condition along both directions is used, the reflection symmetries on the $xz$ -plane $M_y$ and the $yz$-plane $M_x$ are preserved  such that the simple magnetic order in $x$ and $y$-direction are prohibited. Nonetheless, the following composite magnetic order parameter can be used to measure $\mathrm{T}$ breaking in the bulk:
\bea
O_i = \left[c^\dag_i\sigma^zc_i( i c_i^\dag \sigma^x c_{i+x} + i c^\dag_i \sigma^y c_{i+y})+h.c.\right],
\eea
which is odd under time-reversal transformation. This composite magnetic order parameter is the combination of the site spin order in the $z$-direction and the bond spin order in x(y)-direction, which breaks $\mathrm{T}$ but preserves both reflection symmetries in the original Hamiltonian. We then compute the structure factor of this composite magnetic order parameter $S^\text{bulk} = \frac{1}{L^4} \sum_{ij} O_iO_j$, where $i,j$ are over all sites of the bulk, and obtain a finite value after the linear systems size $L$ is extrapolated to infinity, as shown in \Fig{magnetic}(b). This finite-size scaling result indicates when $U>U^\text{bulk}_c$ the system possesses a long-range composite magnetic order and breaks $\mathrm{T}$ in the bulk.

\begin{figure}[t]
\subfigure[]{\includegraphics[height=4.06cm]{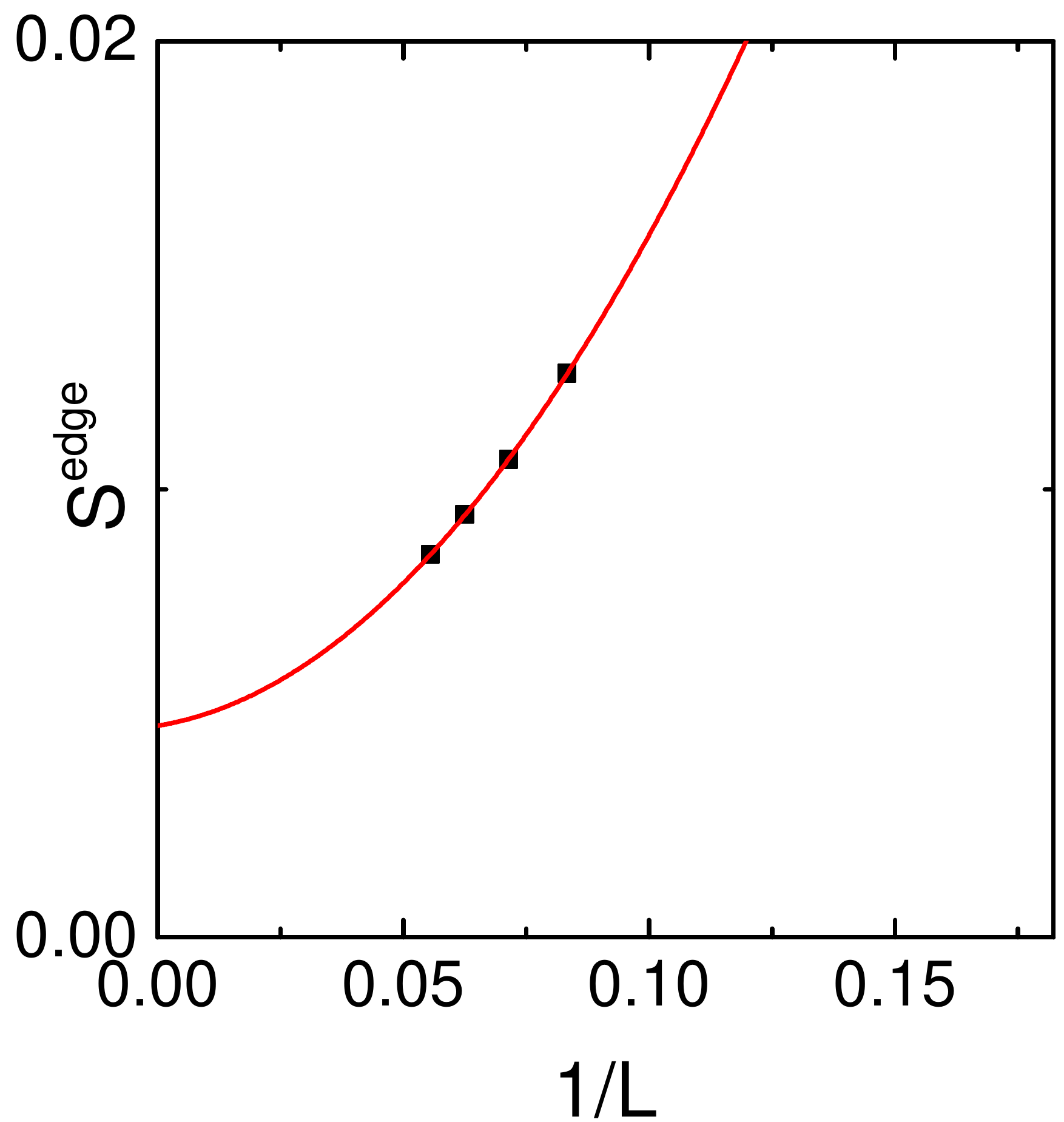}}
\subfigure[]{\includegraphics[height=4.0cm]{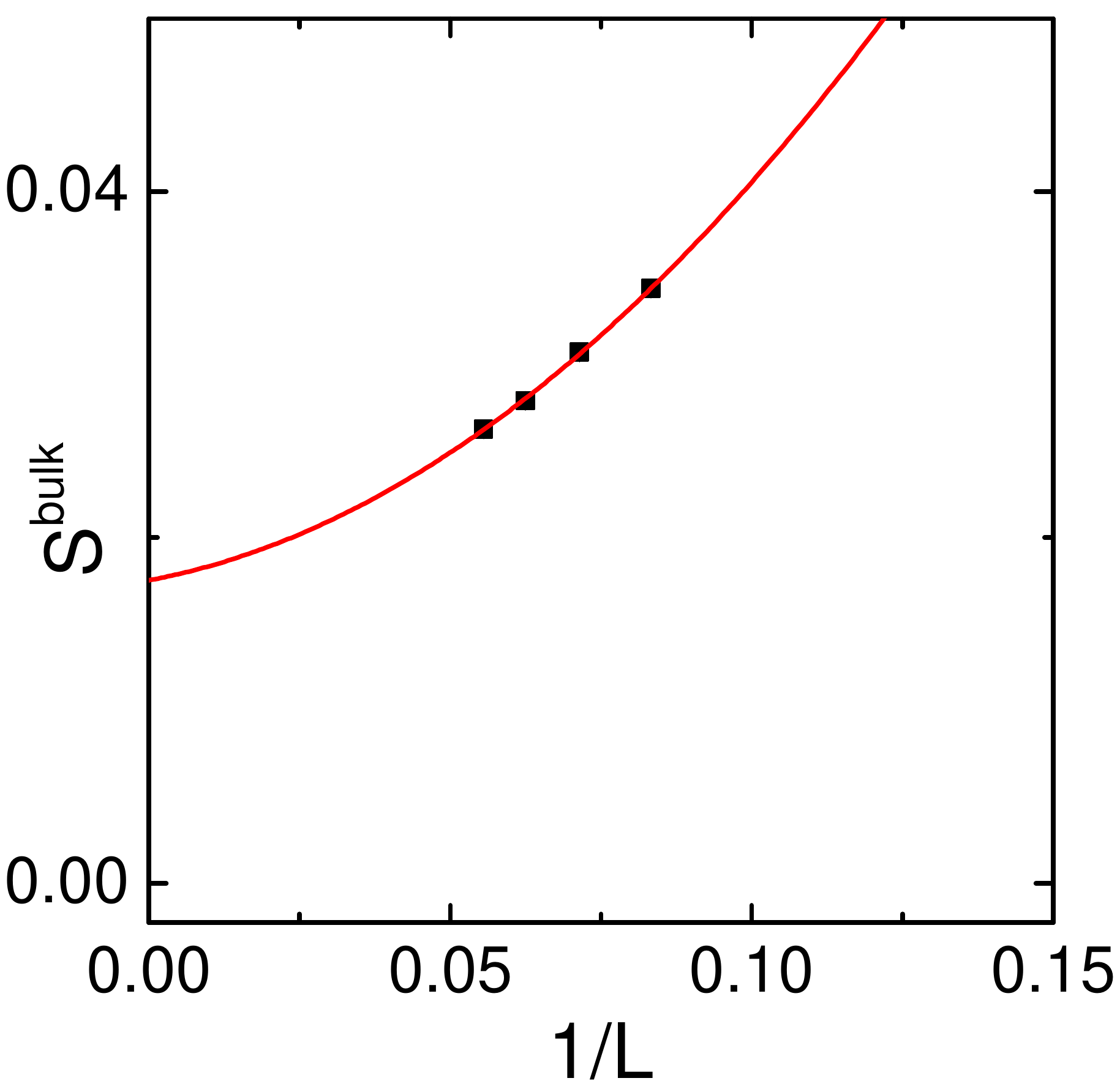}}
\caption{QMC results of the magnetic orders characterizing the breaking of time-reversal symmetry for $\Delta=0.3$. (a) For $U = 4.4>U^\text{edge}_c$, the finite-size scaling of the magnetic structure factor on the edge $S^\text{edge}$ versus $1/L$ shows that a finite magnetic order appears, indicating the breaking of $\mathrm{T}$. (b) For $U = 5.5>U^\text{bulk}_c$, the bulk structure factor of the composite magnetic order $S^\text{bulk}$ versus $1/L$ shows a finite composite magnetic order in the bulk.}
\label{magnetic}
\end{figure}

\begin{figure}[t]
\includegraphics[height=6.0cm]{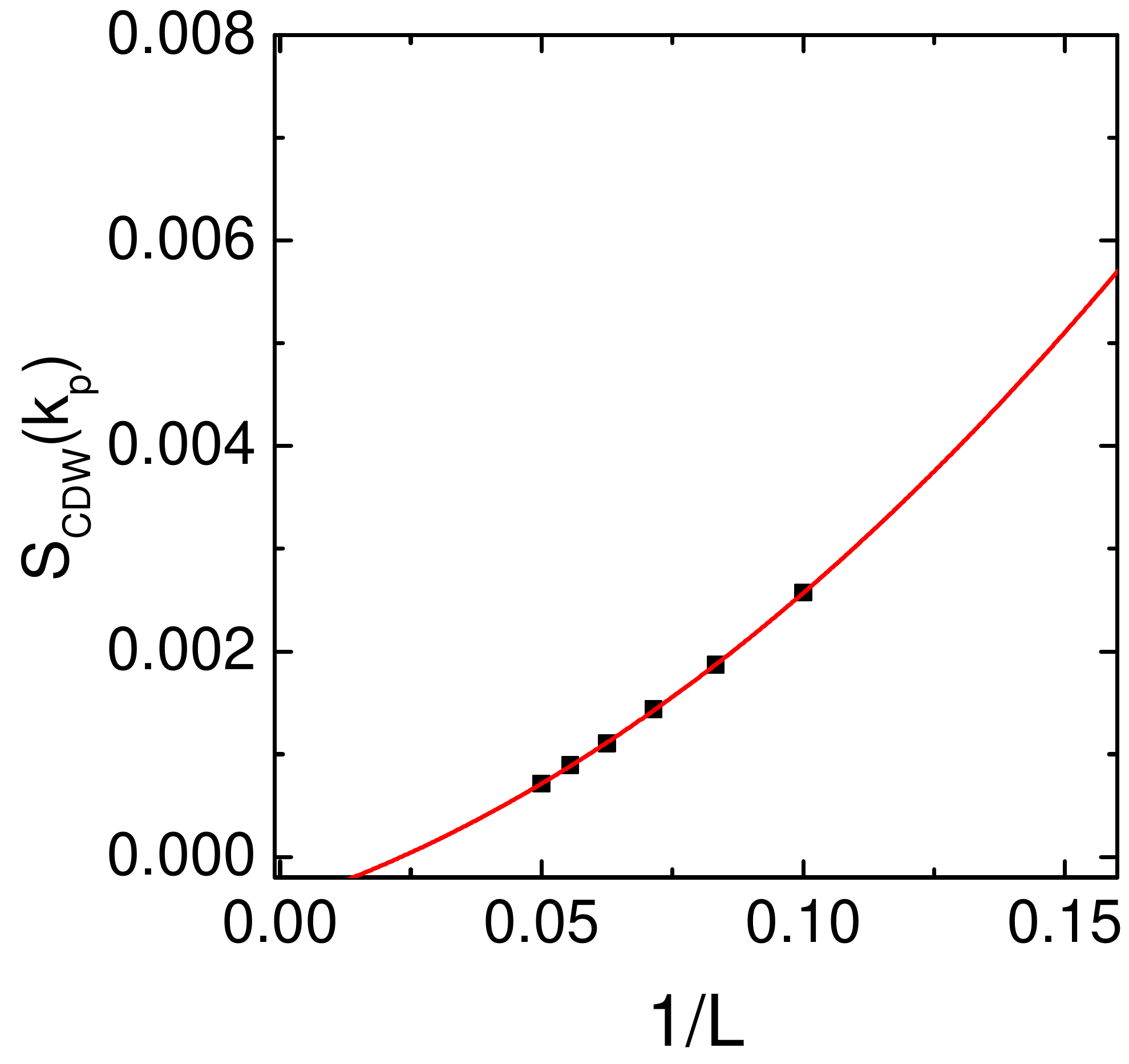}
\caption{QMC results of the dominant CDW structure factor at edge momentum $k_p=\pi$ for $\Delta=0.4$.  For $U = 5.3>U^\text{edge}_c$, the finite-size scaling of edge CDW structure factor $S_{\text{CDW}}(k_p)$ versus $1/L$ shows that CDW order parameter is zero in the thermodynamic limit. This indicates that, in edge time-reversal symmetry breaking phase with s-wave pairing, the CDW order does not appear.}
\label{CDW}
\end{figure}

\subsection{V. Absence of CDW ordering in the T breaking phase}
We investigate the CDW instability in the edge T breaking phase by computing CDW structure factor $S_{\textrm{CDW}}(k) = \frac{1}{L^2}\sum_{ij} n_i n_j e^{i(i-j)k}$ on the edge. The edge CDW structure factor at the peaked momentum $S_{\textrm{CDW}}(k_p)$ are plotted in \Fig{CDW}. The finite-size scaling analysis is performed to extract the CDW order parameter in the thermodynamic limit. The results show that the CDW order parameter vanishes in the thermodynamic limit, which indicates that in the edge T braking phase CDW ordering does not coexist with the s-wave pairing.

\subsection{VI. Critical behaviors at the bulk QCP}
We now study the critical properties of bulk quantum phase transition at $U=U^\text{bulk}_c$ by finite-size scaling. When  $U > U_c^\text{bulk}$, the bulk breaks $\mathrm{T}$ and the transition should belong to the 3D Ising universality class. For both cases of $\Delta = 0.3$  and $\Delta = 0.4$, the data collapse analysis shows that the structure factor can be fitted by a single smooth function, which gives rise to the values of critical exponents: $\eta = 0.04 \pm 0.03 $ and $\nu = 0.62 \pm 0.02 $ for $\Delta = 0.3 $ (see \Fig{bulkcritical}(a)); $\eta= 0.02 \pm 0.03 $ and $\nu = 0.63 \pm 0.03 $ for $\Delta = 0.4$ (see \Fig{bulkcritical}(b)). These results are well consistent with 3D Ising transition.

\end{widetext}

\end{document}